# Optically and electrically excited intermediate electronic states in donor:acceptor based OLEDs

Nikolai Bunzmann[1], Sebastian Weissenseel[1], Liudmila Kudriashova[1], Jeannine Gruene[1], Benjamin Krugmann[1], Juozas Vidas Grazulevicius[2], Andreas Sperlich[1*] and Vladimir Dyakonov[1]

[1] Experimental Physics VI, Julius Maximilian University of Würzburg, 97074 Würzburg, Germany
[2] Department of Polymer Chemistry and Technology at Kaunas University of Technology, Radvilenu pl. 19, LT-50254 Kaunas, Lithuania

(Dated: January 10, 2020)

**Abstract**

Thermally activated delayed fluorescence (TADF) emitters consisting of donor and acceptor molecules are potentially highly interesting for electroluminescence (EL) applications. Their strong fluorescence emission is considered to be due to reverse intersystem crossing (RISC), in which energetically close triplet and singlet charge transfer (CT) states, also called exciplex states, are involved. In order to distinguish between different mechanisms and excited states involved, temperature-dependent spin-sensitive measurements on organic light-emitting diodes (OLEDs) and thin films are essential. In our work we apply continuous wave (cw) and time-resolved (tr) photoluminescence (PL) spectroscopy as well as spin-sensitive EL and PL detected magnetic resonance to films and OLED devices made of three different donor:acceptor combinations. Our results clearly show that triplet exciplex states are formed and contribute to delayed fluorescence (DF) via RISC in both electrically driven OLEDs and optically excited films. In the same sample set we also found molecular triplet excitons, which occurred only in PL experiments under optical excitation and for some material systems only at low temperatures. We conclude that in all investigated molecular systems exciplex states formed at the donor:acceptor interface are responsible for TADF in OLEDs with distinct activation energies. Molecular (local) triplet exciton states are also detectable, but only under optical excitation, while they are not found in OLEDs when excited states are generated electrically. We believe that the weakly bound emissive exciplex states and the strongly bound non-emissive molecular triplet excited states coexist in the TADF emitters, and it is imperative to distinguish between optical and electrical generation paths as they may involve different intermediate excited states.

* sperlich@physik.uni-wuerzburg.de

**I. Introduction**

The major drawback of conventional fluorescent organic light emitting diodes (OLEDs) is that due to spin statistics only 25% of injected electrons and holes form emissive singlet excitons, whereas 75% form long-lived triplets, which mostly decay non-radiatively.[1-4] However, reverse intersystem crossing (RISC) from triplets to singlets is strongly enhanced, if the used materials are designed to exhibit an energy splitting $\Delta E_{ST}$ between the singlet and triplet state in the order of thermal energy $k_B T$.[5-7] In this case, triplets can efficiently be harvested via thermally activated delayed fluorescence (TADF). One approach to achieve a small $\Delta E_{ST}$ is to find appropriate pairs of donor and acceptor molecules, where electrons and holes are located on different molecules to form so-called exciplex states.[7-9] Our work focuses on one of the most prominent donor:acceptor TADF systems which is based on 4,4′,4″-Tris[phenyl(*m*-tolyl)amino]triphenylamine (m-MTDATA) as a donor and Tris(2,4,6-trimethyl-3-(pyridin-3-yl)phenyl)borane (3TPYMB) as an acceptor. After the initial report of efficient TADF from exciplex states between m-MTDATA and 3TPYMB in 2012 by Goushi et al.,[7] several groups dedicated their attention to this system. A variety of experimental techniques were applied in order to rationalize the physics behind the observation of TADF. The used methods include transient electroluminescence (trEL),[10] magnetic field effects,[11,12] diffusion imaging [13] and even photovoltaic studies.[14] However, there is a lack of results from intrinsically spin-sensitive methods. Thus, a detailed understanding of the spin-forbidden upconversion mechanism from triplets to singlets is still missing. In this work we use techniques based on electron paramagnetic resonance (EPR), which are suitable tools to investigate spin states in OLEDs due to their sensitivity to triplets.[15] On the one hand, the broad spectrum of previous results from other experimental approaches allows us to compare our results and assess them within the context of existing models. On the other hand, we are able to fill the gap of missing results from spin-sensitive techniques. While few reports showed the application of transient EPR on films of intramolecular TADF emitters,[16,17] we previously demonstrated the application of EPR based methods on fully processed devices under realistic operation conditions.[18] Here, the methods of electroluminescence and photoluminescence detected magnetic resonance (ELDMR, PLDMR) were introduced and applied to donor:acceptor TADF systems for the first time. In this work, we now provide a more detailed analysis of donor:acceptor systems by application of photophysical characterization methods and a variety of advanced EPR based methods. The sum of the results of those different methods in their entirety allows for a comprehensible interpretation of the magnetic resonance data in terms of a better understanding of TADF emitters. In particular, we address issues regarding the activation energy of delayed fluorescence (DF) and if it is related to the energy gap $\Delta E_{ST}$ between exciplex singlet ($^1$Exc) and exciplex triplet ($^3$Exc) levels, or to molecular triplet states, e.g. to a triplet located on the donor ($^3$LE$_D$) or to a triplet located on the acceptor ($^3$LE$_A$).[19-21] The goal of this work is to reveal, which spin-bearing species are involved in the light generation mechanisms of TADF based OLEDs. In order to make sure that the conclusions we draw from EPR measurements on m-MTDATA:3TPYMB are not restricted to this particular system, we performed additional measurements on further donor:acceptor combinations. We used m-MTDATA as a donor and 4,7-Diphenyl-1,10-phenanthroline (BPhen) as an acceptor as well as Tri(9-hexylcarbazol-3-yl)amine (THCA) [22,23] as a donor with BPhen as an acceptor in order to extend the validity of our findings to a broader spectrum of donor:acceptor systems.

## II. Materials and Devices

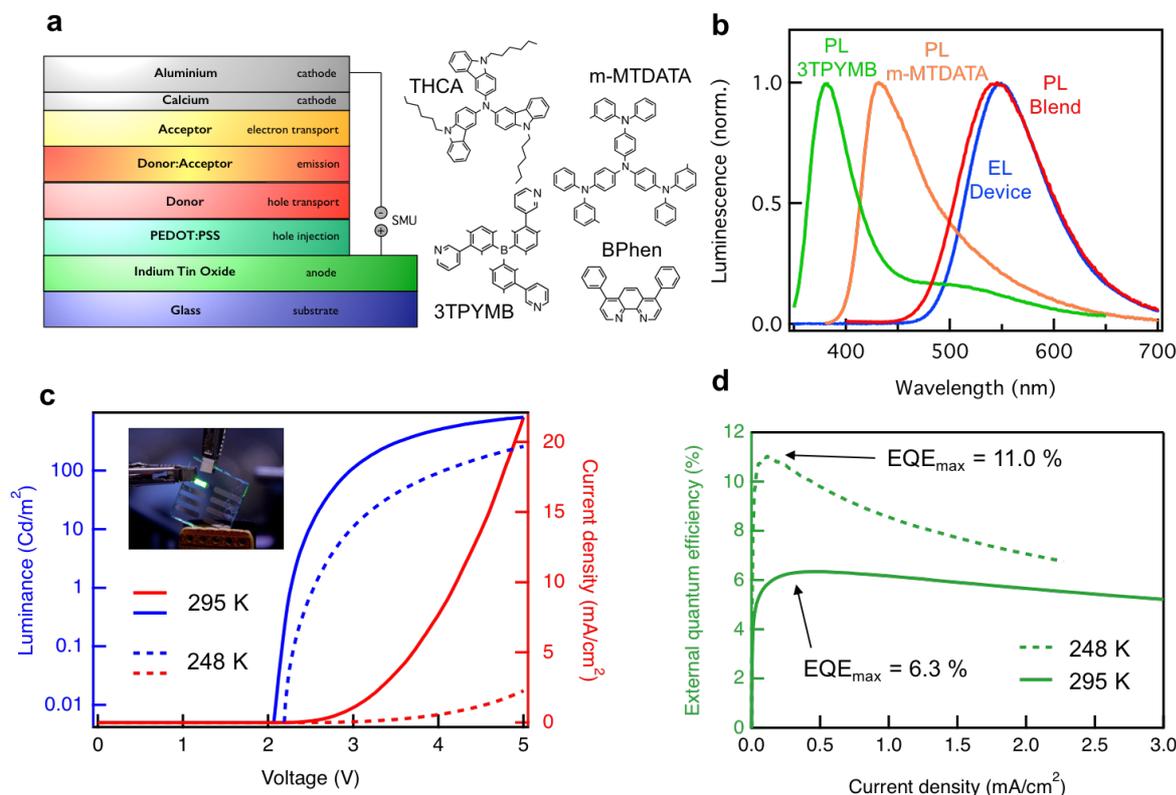

**FIG. 1**. Overview of device layout and electro-optical characterization. **a)** Schematic device structure of donor:acceptor based OLEDs and chemical structures of molecules used in this work **b)** EL and PL spectra of m-MTDATA:3TPYMB together with PL of the pristine materials. **c)** OLED current density and luminance *versus* voltage characteristics of m-MTDATA:3TPYMB at 248 and 295 K. Inset: OLED device under test. **d)** External quantum efficiency *versus* voltage characteristics of m-MTDATA:3TPYMB at T=248 and 295 K.

The molecular components for exciplex based OLEDs investigated in this work are the donor-materials m-MTDATA and THCA combined with the acceptor-materials 3TPYMB and BPhen. **Fig. 1a** shows the corresponding molecular structures and the OLED device structure. An ITO anode covered with a PEDOT:PSS layer is used for hole injection and an aluminum cathode for electron injection. The emission layer consists of a 1:1 mixture of donor and acceptor materials, sandwiched in between layers of the respective pristine molecules, acting either as electron, or hole transport layer. An electroluminescence (EL) spectrum of such a device based on the combination m-MTDATA:3TPYMB as well as a photoluminescence (PL) spectrum of a blended solid film of those molecules and the respective pristine materials are depicted in **Fig. 1b**. One can recognize a clear red shift between the PL of the pure materials and the EL of a device, proving that the emission originates from energetically lower lying exciplex states formed at the interface between the two molecules. The material systems m-MTDATA:BPhen and THCA:BPhen show the same behavior (see **Fig. S2a**, **Fig. S3a**). Exemplary current density and luminance *versus* voltage characteristics for the combination m-MTDATA:3TPYMB at two different temperatures are shown in **Fig. 1c** and the resulting EQE *versus* current density curves in **Fig. 1d**. With the optimized structure which consists of ITO / PEDOT:PSS / m-MTDATA (30 nm) / m-MTDATA:3TPYMB (70 nm, 1:1) / 3TPYMB (30 nm) / LiF (5 nm) / Al (120 nm) we reached a maximum $EQE_{max}$ of 6.3% at room temperature (RT). We measured a photoluminescence quantum yield of 45% in oxygen free m-MTDATA:3TPYMB solid films, which allows for the estimation of a theoretical maximum EQE of 8.3% in OLEDs based on this donor:acceptor system (For details see SI). The maximum EQE we measured for a device comes close to this theoretical value, while the discrepancy between the theoretical and experimental numbers might be attributed to leakage currents. Surprisingly, temperature dependent EQE measurements show that the efficiency increases at temperatures below room temperature. At T=248 K an $EQE_{max}$ of 11.0 % was measured (**Fig. 1d**). While this observation is counterintuitive for TADF, we believe that the suppression of the non-radiative decay at lower temperatures actually outcompetes the decrease of RISC which is why efficiencies can increase below RT.[24] We therefore attribute the limitation of the EQE at RT to non-radiative losses. Still an EQE of 6.3 % at RT exceeds the value of 5% which is the upper limit for purely fluorescent OLEDs. In any case, our spin-sensitive experiments presented in Section IV address the behavior of the reasonably efficient, state-of-the art devices.

## III. Results
### 1. Photoluminescence

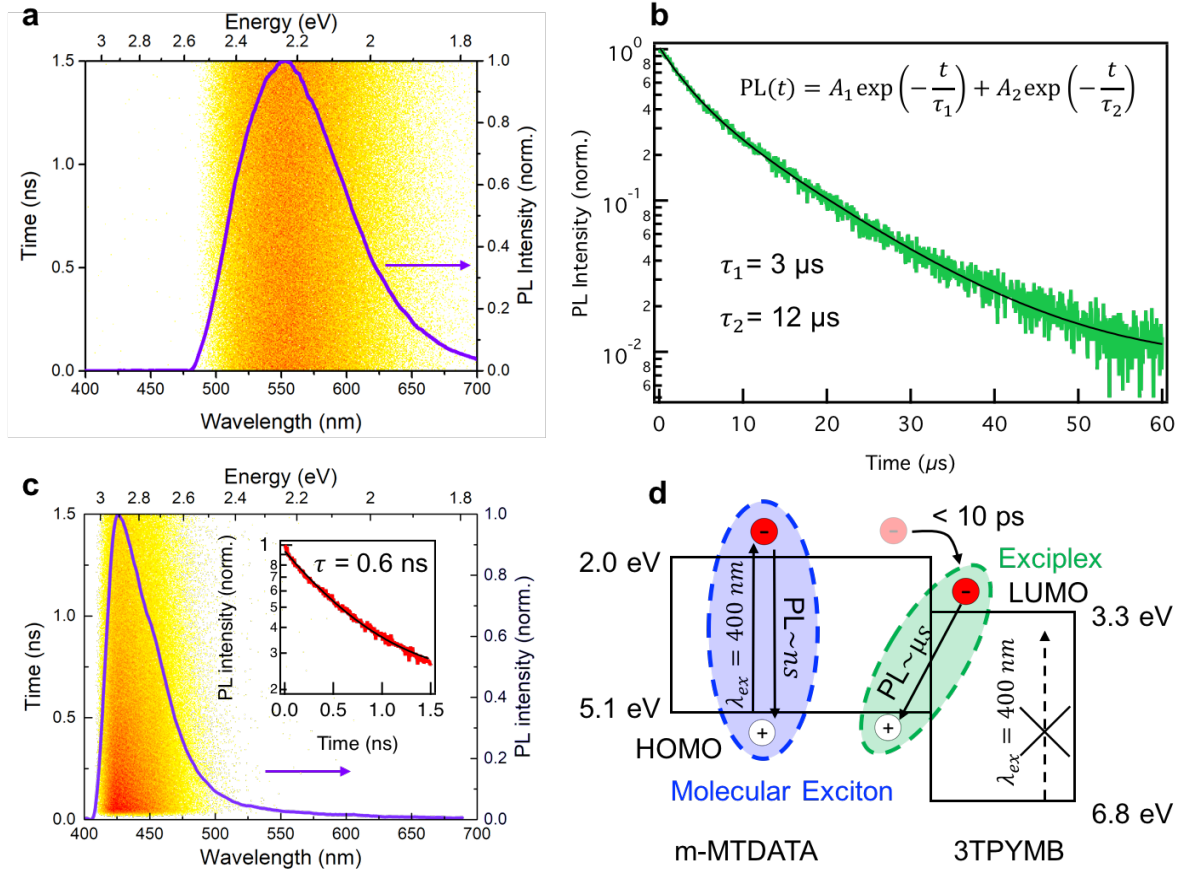

**FIG. 2**. Time-resolved optical spectroscopy. **a)** Streak camera image of PL from an m-MTDATA:3TPYMB blend for excitation at $\lambda_{exc} = 400\ nm$ together with an integrated spectrum over the 1.5 ns detection window. **b)** Transient PL decay of an m-MTDATA:3TPYMB blend. A sum of two exponential decays was used to fit the curve and the corresponding lifetimes were extracted. **c)** Streak camera image of PL from pristine m-MTDATA with an integrated spectrum over the 1.5 ns detection window. Inset: PL decay curve of this emission. A lifetime of 0.6 ns is obtained from fitting the transient with a single exponential decay. **d)** Illustration of the charge transfer process within the instrument response time of 10 ps after optical excitation of m-MTDATA with UV light. Energies for HOMO and LUMO are taken from [7].

In order to further investigate the optical properties of the materials used in this work, we performed time-resolved PL (trPL) measurements on an m-MTDATA:3TPYMB blended solid film. **Fig. 2a** shows a streak camera image which displays emission between 0 and 1.5 ns after excitation. Only emission from the exciplex singlet state ($^1$Exc) at around 550 nm can be observed within the measured time window. **Fig. 2b** shows the PL transient of the observed exciplex emission on a longer time scale proving that the lifetime of the decay is in the microsecond range because of thermally activated fluorescence. Fitting the curve with a double exponential decay yields lifetimes of $\tau_1 = 3\ \mu s$ and $\tau_2 = 12\ \mu s$. Since separated features of prompt and delayed fluorescence cannot be identified in the transient, these lifetimes do not necessarily represent the decay times of prompt and delayed fluorescence, but they demonstrate a distinct prolongation of the overall fluorescence lifetime possibly due to thermally activated RISC. For comparison, a streak camera image of a pristine m-MTDATA solid film was recorded (**Fig. 2c**). Here an emission between 400 nm and 450 nm occurs, reaching its peak at about 425 nm, which corresponds to the cw PL spectrum of m-MTDATA (see **Fig. 1b**). The decay of this emission is presented in the inset of **Fig. 2c.** A lifetime of 0.6 ns is obtained from fitting the transient with a single exponential decay. Considering the negligibly small absorption of the exciplex state itself [21,25,26] it is remarkable that only exciplex emission is observed in the blended film while emission from m-MTDATA is completely quenched. According to photoexcitation spectra (**Fig. S1**), with the 400 nm laser in the streak camera measurement only m-MTDATA can be excited in the blended film**.** We conclude that the initial singlet excitation ($^1LE_D$) of the donor molecule m-MTDATA is followed by an electron transfer to the acceptor molecule 3TPYMB and the formation of an exciplex state within the instrument response time of 10 ps. This ultra-fast electron transfer outcompetes PL from m-MTDATA, which is similar to what is usually observed in donor:acceptor blends for organic photovoltaics.[27,28] Intersystem crossing (ISC) from singlet ($^1LE_D$) to triplet state ($^3LE_D$) can also contribute to the efficient quenching of the m-MTDATA emission and will be discussed on the basis of magnetic resonance data in more details below. An overview of the involved processes is schematically shown in **Fig. 2d**, illustrating how the exciplex state is formed after the charge transfer process, giving rise to ~µs long-living emission at 550 nm. For the sake of completeness, streak camera measurements were also performed on blended solid films of m-MTDATA:BPhen and THCA:BPhen (see **Fig. S2b**, **Fig. S3b**). For m-MTDATA:BPhen, exciplex singlet emission at 560 nm can be observed within the measured time window of 500 ps. In contrast to the m-MTDATA:3TPYMB blend, there is additional emission between 410 and 430 nm at early times, which can be assigned to m-MTDATA. However, this emission decays within approximately 10 ps, which is much faster than the decay of pristine m-MTDATA, as shown in **Fig. 2c**. Again, the time scale, on which the CT from m-MTDATA to BPhen and the formation of the exciplex state take place, remains very short. The PL decay of an m-MTDATA:BPhen blended film on a longer time scale is shown in **Fig. S2c**, where the ~µs long-lived emission of the exciplex becomes apparent. Here, a sum of two exponential decays was not sufficient to reasonably fit the curve and instead a sum of two stretched exponential decays was used. Stretched exponentials apply when a distribution of lifetimes is given, which is reasonable for donor:acceptor systems where the molecules are randomly oriented and the wave-functions overlap determining $\Delta E_{ST}$ can vary.[29,30] We extracted lifetimes of $\tau_1$ = 30 ns and $\tau_2$ = 0.2 µs (More details about this fitting procedure are given in **Fig. S2c**). The excitation of BPhen with the 400 nm wavelength in a streak camera experiment was not possible according to photoexcitation spectrum shown in **Fig. S2a,** which explains the absence of any emission from BPhen. The streak image for a THCA:BPhen blended solid film shows exciplex singlet emission at around 560 nm within the measured time window of 1.5 ns and a distinct signature of THCA emission at around 475 nm. Since the emission from THCA is also visible in the steady-state PL from the THCA:BPhen blended solid film, it is likely that the blend ratio for the measured sample deviates from 1:1. A comparison with the PL decay of a pristine THCA solid film (**Fig. S3d**) shows that the PL lifetime of THCA is considerably shortened in the blend with BPhen proving the occurrence of CT between THCA and BPhen. The PL decay of THCA:BPhen (**Fig. S3c**) demonstrates ~µs long-lived emission of the resulting exciplex state. Here fitting the curve with a sum of two stretched exponential decays yields lifetimes of $\tau_1$ = 13 ns and $\tau_2$ = 3 µs. In summary, all three donor:acceptor combinations exhibit similar photophysics, in which optical excitation of the donor molecule is followed by a fast charge transfer to the acceptor molecule upon formation of a long-lived emissive exciplex state.

## 2. Magnetic Resonance

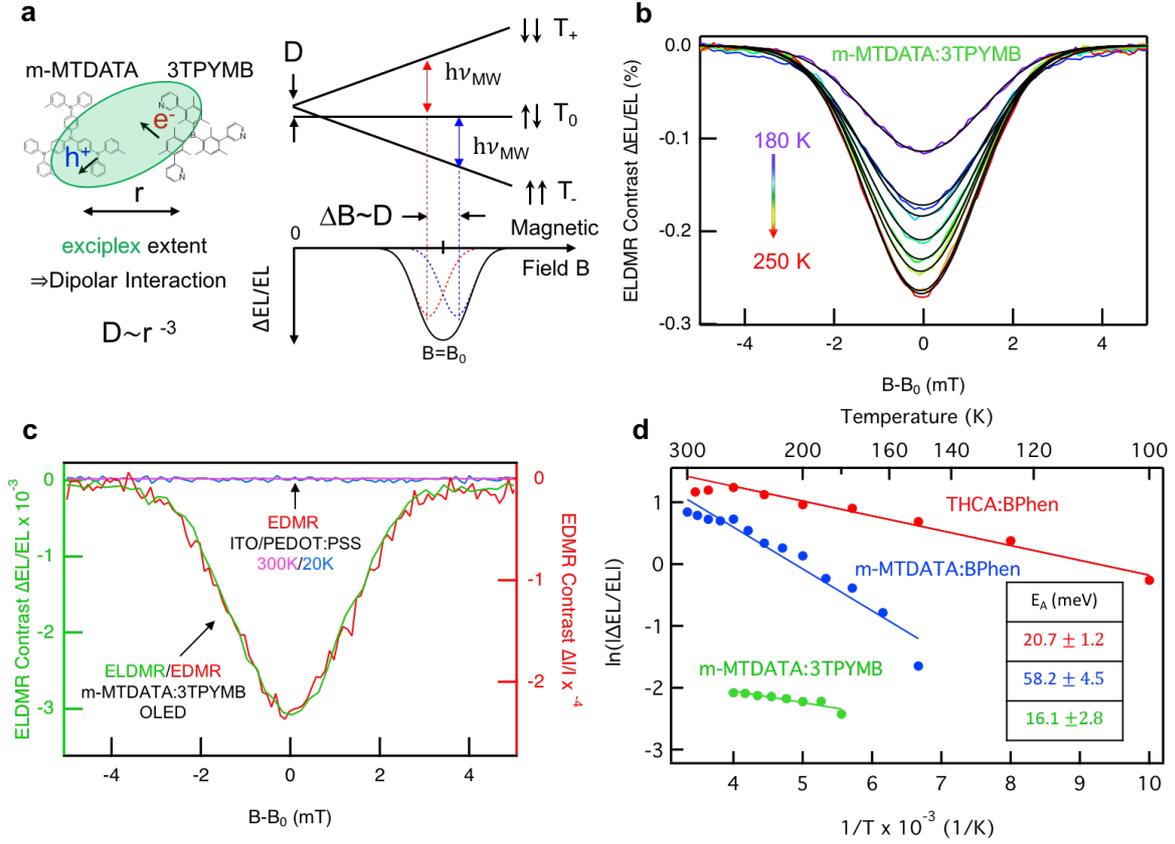

**FIG. 3**. Temperature dependent spin-resonance experiments on OLEDs. **a)** Illustration of exciplex states formed at the donor-acceptor interface and EL change due to spin-flip transitions between exciplex triplet states in an external magnetic field *B*. The exciplex wave function extent r can be understood as spatial separation of hole and electron spins with magnetic dipolar interaction $D \sim r^{-3}$. Spin-flip transitions between Zeeman levels are induced by microwave photons ($h\nu_{MW}$) and can be detected via relative EL change ΔEL/EL (or current ΔI/I) as a Gaussian bell-shaped curve centred at $B=B_0$ with the width ΔB determined by *D*. **b)** Temperature-dependent EL contrast (ΔEL/EL) under magnetic resonance conditions (ELDMR) for m-MTDATA:3TPYMB together with Gaussian fits (black lines). ELDMR spectra for other studied material systems are shown in **Fig. S7**. **c)** ELDMR and EDMR (ΔI/I) spectra of m-MTDATA:3TPYMB are identical in shape implying the same spin-states are involved in recombination. As a reference, a PEDOT:PSS-only device (without emitter layer) shows no spin-resonance effect on the current, which excludes spin-dependent effects in the transport layers. **d)** Arrhenius plot of the integrated ELDMR spectra. The activation energy $E_A$ is derived from the slope of linear fits and tabulated in the inset for three studied material systems.

In the next step, we applied EPR to OLEDs. Here microwaves of a fixed frequency $\nu_{MW}$ (in the GHz range) drive transitions between Zeeman sublevels of a triplet spin state which is energetically split due to an external magnetic field *B*, as schematically shown in **Fig. 3a**. As soon as the following resonance condition is fulfilled, microwave-induced transitions take place:

$$h\nu_{MW} = g\mu_B B \Delta m_s \pm D\left(\cos^2\theta - \frac{1}{3}\right) \quad (1)$$

Here *h* is the Planck constant, *g* is the g-factor of the spin system, $\mu_B$ is the Bohr magneton, $\Delta m_s = 1$ is the allowed change of the magnetic quantum number, *D* is the magnetic dipole-dipole interaction of two spins and $\theta$ is the angle between the direction of the external magnetic field and the vector connecting the two spins. For a statistical distribution of molecular orientations and electron-hole separations in heterogeneously mixed blends, a distribution of magnetic dipolar interaction energies $D$ can be expected. This results in a superposition of several possible EPR transitions. If $D$ is small, as in the case of distant electron-hole pairs, the EPR spectrum for an S=1 (triplet) state will consist of a single inhomogeneously broadened Gaussian bell-shaped curve. Alternatively, for strongly interacting close-by pairs of spins (large $D$), as for molecular (localized) triplet excitons, a broad spectrum ("powder pattern") with pronounced, separated side-peaks and shoulders is expected.[31,32]

In contrast to classical EPR, where microwave absorption in resonance is measured, we probe parameters which are directly related to optoelectronic properties of our samples, namely the EL or current I in OLEDs, or PL in films. This creates a direct link between the spin species and the recombination or transport processes in devices or films. The respective techniques are called EL-detected magnetic resonance (ELDMR), electrically detected magnetic resonance (EDMR) and PL-detected magnetic resonance (PLDMR). In ELDMR and EDMR, the operating device is electrically driven, whereas in PLDMR, the donor:acceptor film is optically excited. Spin-flip transitions are particularly efficient, if the population difference between Zeeman sublevels is large enough (i.e. non-Boltzmann, 0.15% at RT). There are several mechanisms for this, either considering transformation between singlet and triplet excitons, or between singlet and triplet exciplex states. In neutral excitations, such as triplet states, the polarization can be achieved via spin-selective population of a particular Zeeman sublevel, e.g. of $m_S$=0 in an optical pumping cycle via ISC from initially photogenerated singlet excitons.[15] Alternatively, selective depopulation of a particular Zeeman sublevel can also lead to polarization. In the case of weakly bound electron-hole pairs (CT or exciplex states, as in our case), optical excitation leads to formation of singlet excitons with 100% yield. If CT takes place, only singlet exciplex states can be formed due to spin conservation rule. Since the electron-hole dipolar coupling is weak (distant pairs), singlet and triplet exciplex states are energetically very close (almost degenerate). Therefore, a singlet-triplet mixing can occur, either via hyperfine interaction or via a so-called Δg-mechanism, depending on the material systems,[12,16] which in turn can lead to spin polarization of the triplet exciplex state, but only if the external magnetic field lifts the degeneracy of the Zeeman sublevels. In contrast to optical generation, electrical injection of charges leads to the statistical formation of electron-hole pairs with 75% share of triplets. If the rate constants for non-radiative decay and RISC are different between the three triplet Zeeman sublevels, a non-Boltzmann distribution occurs. In other words, under optical and electrical excitations, a steady-state population difference between the spin states can build up, and microwave-induced spin-flip transitions will change recombination rates. This results in a change of EL, I or PL depending on the applied method. We note that an additional relaxation pathway may open if the energy of a molecular triplet exciton state is lower than the energy of the triplet exciplex state. Such a spin-conserving process can form triplet excitons from triplet CT or exciplex states.

To clarify the spin sensitive mechanisms in OLEDs, ELDMR is the most suitable method as it directly probes EL. Temperature-dependent ELDMR spectra measured on an m-MTDATA:3TPYMB OLED are shown in **Fig. 3b**. Each spectrum consists of a single, Gaussian shaped line centered at the magnetic field corresponding to g=2.002 in **Eq.(1)**. The g-factor is close to the free-electron value and typical for radicals, but also for triplet states in organic semiconductors due to the small spin-orbit coupling in carbon based materials, free of heavy atoms.[15,33] Remarkable is the relative change ΔEL/EL in resonance of almost 1%, which is large compared to the population difference between Zeeman sublevels expected from Boltzmann statistics (0.3-0.15% at 150-300K). According to **Eq.(1)** it is difficult to distinguish between S=1/2 doublets and S=1 triplets when D is small, e.g. in weakly interacting spin pairs, since two triplet transitions overlap and appear as one envelope curve (**Fig. 3a**). Further, although we probe electron-hole recombination from the excited singlet exciplex state (EL), we can only manipulate the population of the triplet Zeeman sublevels. To understand the origin of the observed magnetic resonance effect on EL, i.e. to explain why and how spin-flip transitions in a non-emissive triplet exciplex state ($^3$Exc) lead to a change of EL from the singlet exciplex state ($^1$Exc), we need to know the sign of the EL change. The lock-in based ELDMR technique does not allow to do this unambiguously. Therefore, we independently measured the EL intensity directly with a photodiode connected to a high-sensitivity digital oscilloscope, while applying the resonant on-off microwave pulses (**Fig. S4**). By doing so, we clearly observed a decrease in EL (negative sign) and correspondingly plotted the ELDMR spectra, as shown in **Fig. 3b**. Independently of this, we measured the magnetic field effect on EL [12,34] and found that EL increases with magnetic field (**Fig. S5**), which is due to the fact that the magnetic field lifts the degeneracy of triplet sub-levels (Zeeman effect). This behavior is in agreement to the negative sign of ELDMR since the resonant transitions connect the inner (strongly occupied) and outer Zeeman sublevels and accelerate the non-radiative recombination via triplet channel. Remarkably, the magnitude of the ELDMR contrast decreases with decreasing temperature. This is completely opposite to what is commonly observed in EPR experiments, where lower temperatures lead to an increase in spin polarization according to Boltzmann statistics and therefore enhanced signals.[15] ELDMR spectra for OLEDs based on m-MTDATA:BPhen and THCA:BPhen show similar line shapes and temperature dependencies (**Fig. S7a, b**). In addition, the shape of the ELDMR spectra is independent of whether OLEDs are processed from the solution or evaporated in a vacuum (**Fig. S7c**). To ensure the same exciton generation rate for each temperature, all ELDMR spectra where measured at the same current density of 1 mA/cm$^2$. Consequently, the observed temperature dependence can be attributed to the thermal activation of RISC in TADF-based OLEDs, unambiguously proving the TADF nature of the observed EL. From these experiments we can conclude that electrically generated triplet exciplex states are spin-polarized already at room temperature, since lifetimes of $m_S=\pm 1$ states and $m_S=0$ states differ, which is therefore responsible for a non-Boltzmann distribution. The latter $m_S=0$ state is linked to the singlet exciplex state via RISC and therefore kinetically controls the EL intensity. The details of the RISC mechanism are also non-trivial. Recent reports attribute the driving force of this RISC process in donor:acceptor-based TADF emitters to the so-called $\Delta g$-mechanism.[12,34] This mechanism facilitates intersystem crossing between the singlet state S and the triplet state $T_0$ due to a difference in spin precession frequencies $\Delta \omega_P$ which arises from a difference in electron spin g-factors $\Delta g$ in the presence of an external magnetic field $B$ ($\Delta \omega_P = \mu_B \Delta g B/\hbar$).[35] For donor:acceptor systems, a contribution of the $\Delta g$-mechanism can be expected because electron and hole, which form the exciplex state, reside on adjacent non-identical molecules. In order to obtain an estimate value for $\Delta g$ in our OLEDs we performed ELDMR measurements over an extensive range of magnetic fields (25 mT – 1.4 T) and microwave frequencies (0.7 GHz – 38 GHz) (**Fig. S6**). Here, an increase of the linewidth with increasing frequency is observed which indicates a non-negligible $\Delta g$. We used the software package EasySpin [36] to perform a global fit of all spectra in the accessible frequency range and extracted an upper limit for $\Delta g$ of $9.2 \cdot 10^{-4}$. We emphasize that this value is derived directly from spectroscopic data and is not deduced by modelling magnetic field effects, such as magneto electroluminescence (MEL). In [12], the assumption of $\Delta g = 10^{-4}$ was sufficiently large to explain magnetic field effects in m-MTDATA:3TPYMB based OLEDs with a dominant $\Delta g$-mechanism. Investigations on other organic materials report values for $\Delta g$ in the range of $10^{-3}$ to $10^{-4}$.[37] Our measurement is consistent with these numbers and thus supports the scenario that $\Delta g$-mechanism is responsible for RISC.

To ensure that the observed ELDMR signals are directly related to the emitting layer, but not to the spin-dependent transport or injection in the adjacent transport layers or at the interfaces, comparative EDMR measurements on fully processed OLEDs without TADF emissive layers were performed. To remind, EDMR probes the microwave-induced change of current through the OLED while a constant voltage is applied to the device. ELDMR and EDMR on an OLED (m-MTDATA:3TPYMB) yield identical signals, as shown in **Fig. 3c** (see also **Fig. S8** for THCA:BPhen). However, the relative change of the current in resonance is about one order of magnitude smaller than the EL contrast. The reference sample, consisting of ITO/PEDOT:PSS without donor and acceptor layers, yields no EDMR signal at temperatures between 20 K and 300 K at all. These observations clearly show that the ELDMR and EDMR signals originate from the same spin-dependent mechanisms in the emissive layer and we can exclude spin-dependent transport, or injection in the anode and cathode layers as source of the observed effects. Moreover, we speculate that the EDMR signal can actually be induced by the ELDMR effect. A change of the recombination rate of exciplex states in magnetic resonance under constant voltage conditions can give rise to a change in the current. Consequently, the occurrence of an ELDMR signal can induce a smaller EDMR signal of identical shape. Similar correlations between a change in EL and current are observed in magnetic field effect studies on donor:acceptor based TADF OLEDs.[12] We additionally measured ELDMR and EDMR spectra at different driving currents and found that both parameters are dependent on the current density through the device (**Fig. S9a**). Increasing ELDMR contrast with increasing current may be a signature of polaron- exciplex interaction, as also known for other types of OLEDs.[33,38]

From the temperature dependence of the ELDMR signal intensities we can calculate the activation energy $E_A$ from an Arrhenius plot. ELDMR is not yet a widely used method to determine $E_A$, although it probes the OLED response to a very fast spin-flip in the electron-hole pair, i.e. without possible artefacts due to charge injection, transport and exciplex formation. **Fig. 3d** shows Arrhenius plots of the integrated ELDMR spectral intensity for OLEDs made with three different material systems. From the slope of linear fits, values for $E_A$ were calculated for each material system. These values are in the range between 16 and 58 meV, which is in good agreement with the values for $\Delta E_{ST}$ of other donor:acceptor-based TADF systems.[7,39,40] Based on the assumption that the singlet exciplex is the emissive state and its rate-limited (de-)population occurs via RISC from the triplet exciplex state, we consider the deduced activation energies as good approximations for the singlet-triplet gap $\Delta E_{ST}$, since ELDMR probes thermally activated spin-conversion.

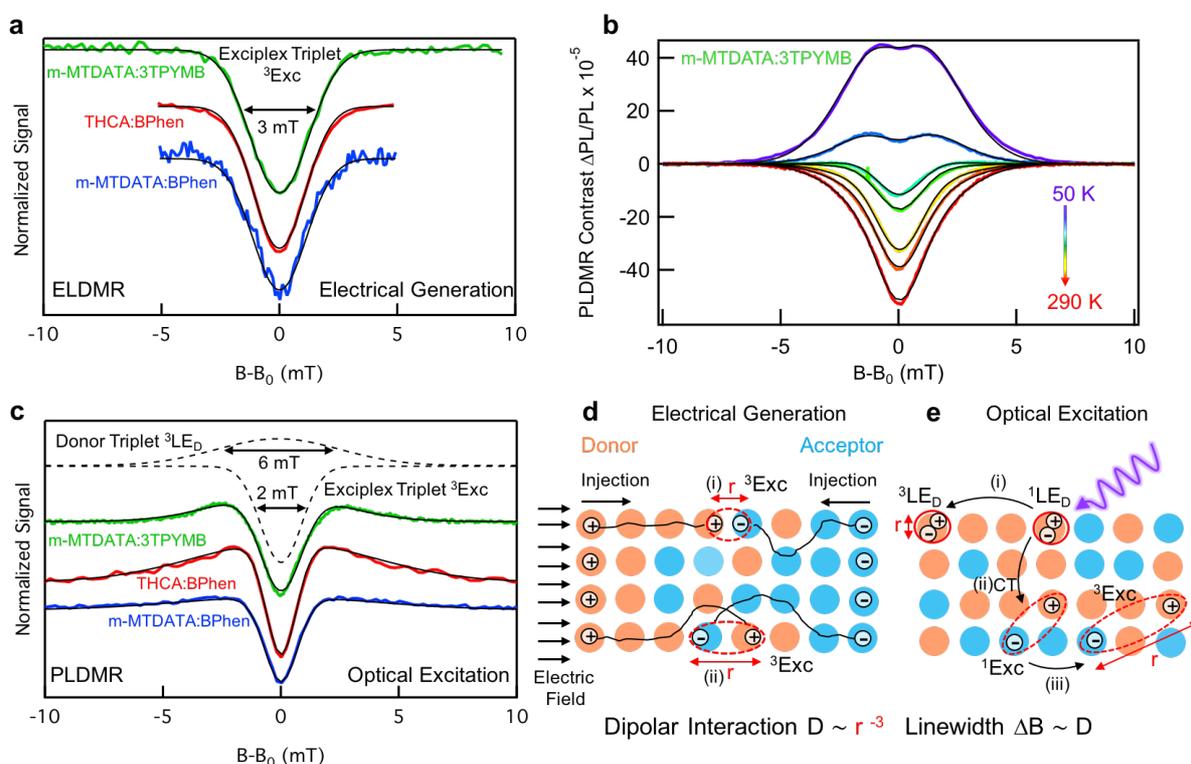

**FIG. 4**. Comparison of ELDMR and PLDMR spectra. **a)** Normalized ELDMR spectra for different donor:acceptor combinations. All systems exhibit a single Gaussian line assigned to the exciplex triplet ($^3$Exc). The shown curves are each recorded at T=200 K. **b)** Temperature dependent PLDMR spectra of an m-MTDATA:3TPYMB solid film. Each spectrum consists of a broad and a narrow Gaussian line, which are superimposed. **c)** Comparison of normalized PLDMR spectra for different donor:acceptor systems, all recorded at T=200 K. The dashed line shows separate fits of the broad and narrow components for m-MTDATA:3TPYMB (green line). We assign the narrow component to the exciple triplet ($^3$Exc) and the broad component to a molecular triplet on m-MTDATA ($^3LE_D$). **d)** Illustration of the triplet formation after the electrical injection of charge carriers. Electrons and holes are injected into the mixed layer of the OLED and form excited states, 75% of which are triplet exciplex states ($^3$Exc). The spatial separation of the e-h pairs r can vary by the electric field. **e)** Illustration of the triplet formation under optical excitation. Initially, only singlet excitons are formed in the donor phase ($^1LE_D$). They can diffuse and undergo ISC to form molecular triplet excitons ($^3LE_D$) (i). Alternatively, the donor singlet ($^1LE_D$) can undergo a charge transfer to form an exciplex singlet ($^1$Exc) (ii). Subsequently, an exciplex triplet ($^3$Exc) can be formed, e.g. via the $\Delta g$-mechanism (see text) or another spin-conversion process (iii). Note that the dipolar-coupling $D$ and hence the ELDMR/PLDMR line width $\Delta B$ can vary due to the broad distribution of the electron-hole separations r.

In order to understand the radiative recombination pathway in more detail, we now compare the spin-dependent recombination of exciplex states formed by electrical injection with photogenerated exciplex states. For this we first compare normalized ELDMR spectra for three different donor:acceptor combinations as shown in **Fig. 4a**. Although, we clearly see a difference in activation energies of the underlying RISC process in these blends, the ELDMR spectra are almost identical for all studied OLEDs, with the same resonance position (i.e., in the vicinity of g=2.002), similar signal shape and only slightly different linewidth of about 3 mT FWHM (full width at half maximum). This similarity can be explained by the fact that the donor molecule in each case is either THCA or m-MTDATA, which have very similar molecular structures, as shown in **Fig. 1a**. Although the electronic molecular environment determines the g-factor and hence the structure of the ELDMR signal, g-factor differences are too small to be resolved in the used frequency range. On the other hand, an exciplex is a delocalized electron-hole pair over donor and acceptor molecules. As we will discuss in more detail later, the spatial separation between them is the parameter that might determine the ELDMR linewidth and since we expect a broad distribution of such pairs in the emissive blends, the influence of the structure of the involved individual molecules appears to be small and ELDMR signals appear rather similar for several donor:acceptor combinations.

Next, we measured PLDMR on m-MTDATA:3TPYMB blended solid films at temperatures between 50 and 290 K, as shown in **Fig. 4b**. The first thing one notices is another shape of the spectrum. All spectra consist of a superposition of a narrow and a broad component with Gaussian shape and the broad component becomes more and more visible as temperature decreases. As already mentioned, the sign determination with the lock-in amplifier is ambiguous, especially if there is a phase shift between individual spectral components which is temperature-dependent, as in our case.[41,42] Also here we used an oscilloscope for direct PL detection with pulsed resonant microwaves instead of a lock-in amplifier (**Fig. S4**). The broad component exhibits a positive sign and the narrow component a negative one. PLDMR spectra from blended solid films of the donor:acceptor combinations m-MTDATA:BPhen and THCA:BPhen are shown in **Fig. S10** and exhibit a similar behavior. Remarkably, the broad components in these blends are only pronounced at low temperature, whereas at room temperature the narrow signal dominates strongly. **Fig. 4c** presents normalized PLDMR spectra of all studied donor:acceptor blends showing that each system exhibits the same behavior. In order to separate the two contributions, we exemplary fitted the PLDMR spectrum of m-MTDATA:3TPYMB with two Gaussians, shown as dashed lines in the upper part of **Fig. 4c**. The narrow PLDMR component is very similar to the ELDMR signals in sign and form, but has a slightly smaller linewidth of 2 mT compared to 3 mT in ELDMR. The reason for this can be slightly different spatial distributions of electron-hole pairs, as illustrated in **Fig. 4d,e**. But the broad PLDMR component with a linewidth of 6 mT is clearly a new feature not present in ELDMR. In order to exclude the influence of the film preparation method, we performed PLDMR on evaporated and solvent-processed solid films with the same outcome (**Fig. S11**).

## V. Discussion

In order to assign the detected signals in ELDMR and PLDMR to particular excited states, we assume that electrical injection predominantly populates the energetically lowest triplet state according to spin statistics. In the donor:acceptor blends studied in this work this is the triplet exciplex state ($^3$Exc) (Energies of all molecular singlet and triplet states as well as exciplex singlet states are shown in **Fig. S12**). The unusual temperature behavior strongly supports the scenario that the signals occurring in ELDMR and EDMR are due to thermally-activated RISC between triplet ($^3$Exc) and singlet exciplex ($^1$Exc) states. Although the narrow components in PLDMR and in ELDMR have slightly different widths, they are very similar in shape and, most importantly, in temperature-dependence. Directly measured microwave pulse-induced changes of PL (**Fig. S4b**) and EL (**Fig. S4d**) show that the narrow PLDMR and ELDMR signals show up themselves as transients crossing the reference baseline (off-resonance measurement). These observations demonstrate that they are of the same origin and we assign the narrow PLDMR peak to exciplex triplets ($^3$Exc), too. To further verify that the narrow PLDMR peak originates from the exciplex triplet, we tested if this assignment is consistent with excitation power dependent PLDMR measurements (**Fig. S9c**). Higher order processes such as triplet-triplet annihilation (TTA), which involve more than one exciton, would show an excitation dependent amplitude of the PLDMR contrast ΔPL/PL because such processes depend on the density of excitons. We however observe an excitation intensity independent amplitude of the PLDMR contrast. This is consistent with the narrow PLDMR component originating from RISC where only a single exciplex triplet is needed for creation of an emissive singlet. Although both PLDMR and ELDMR probe the exciplex triplet, their relative magnitudes (ΔPL/PL and ΔEL/EL) differ by at least a factor of 10. This can be due to different spin statistics. For optical excitation, triplet exciplex states are formed via ISC from optically generated singlets whereas for electrical injection triplet exciplex states are directly formed with a probability of 75% thus giving rise to a higher ELDMR contrast.

The second component in the PLDMR signal is a factor of two broader and therefore should have a different origin. We tentatively attribute it to molecular triplet excitons ($^3$LE), either on the donor or the acceptor molecule. The more localized character leads to a stronger dipolar-coupling $D$ than in exciplex triplets, which are delocalized over at least two molecules. As schematically shown in **Fig. 3a**, $D$ determines the linewidth of these magnetic resonance spectra ΔB. Therefore, PLDMR from a localized triplet exciton on the donor or acceptor molecule is expected to be broader. Remarkably, PLDMR measurements on blended solid films of m-MTDATA:3TPYMB and m-MTDATA:BPhen exhibit an additional information, a so-called half-field signal (**Fig. S13**). These signals arise from a spin-flip transition with a change of spin quantum number $\Delta m_s = 2$, i.e. between $T_+$ and $T_-$ Zeeman states and are detected at half the magnetic field $B_0/2$ of the full-field transition centered at $B=B_0$. Their occurrence is unambiguous proof that the signal stems from a high spin state, since spin ½ particles cannot show a $\Delta m_s = 2$ transition. On the one hand that observation excludes polarons (spin ½) as the origin of the PLDMR signal and on the other hand it is a strong hint for the involvement of a local triplet. The intensity of the half-field transition is proportional to $r^{-6}$ where r is the distance between the spin carrying particles.[43] Since r is expected to be relatively large for exciplex triplets, the occurrence of a half field signal is not expected for exciplex states. We therefore conclude that the half-field signal must be assigned to a local triplet where r is expected to be smaller.

An important consequence of the above assignment is that our experiments do not provide any evidence for the involvement of local triplets ($^3LE_D$ or $^3LE_A$) in the emergence of delayed fluorescence in electrically driven devices although our detection scheme is sensitive enough to probe them. Therefore, we cannot justify a scenario, in which the exciplex state couples to energetically higher lying local triplets via spin-orbit coupling,[19-21] as in this case we would expect triplet signatures in ELDMR, too.

We now discuss the relationship between the width of the ELDMR/PLDMR lines and the spatial separation of electron and hole, which form a bound state. From EPR spectra, the distance between electron and hole $r_{e-h}$, can be estimated [44] by the following equation:

$$r_{e-h}[nm] = \sqrt[3]{\frac{2.785}{D[mT]}} nm \qquad (2)$$

Here, $r_{e-h}$ is obtained in units of nm, if $D$ is used in units of mT. We use FWHM as an *upper limit* for $2D$, as we cannot fully exclude other mechanisms of the EPR line broadening, e.g. unresolved hyperfine interactions with surrounding nuclei. Consequently, a *lower boundary* for $r_{e-h}$ can be calculated by using **Eq. (2)**. From the FWHM of our ELDMR spectra, one finds $2D \leq 3$ mT resulting in $r_{e-h} \geq 1.2$ nm (electrical generation), while for the narrow PLDMR component we estimate $2D \leq 2$ mT resulting in $r_{e-h} \geq 1.4$ nm (optical excitation). The broad $^3$LE PLDMR component yields $2D \leq 6$ mT resulting in $r_{e-h} \geq 1.0$ nm. These numbers can be explained by the different triplet formation mechanisms for electrical injection and optical excitation. **Fig. 4d** shows an illustration of the triplet exciton formation in the case of electrical generation. Here electrons and holes form exciplex triplets ($^3$Exc) in the emission layer, which consists of a mixture of donor and acceptor molecules (see also **Fig. 1a**). Attar et al. reported that a voltage, which is applied to a mixed m-MTDATA:3TPYMB layer, broadens the distribution of distances between electrons and holes forming exciplex states.[25] Depending on the orientation of the exciplex dipoles with respect to the E-field, the electrostatic force causes either their compression (**Fig. 4d (i)**) or expansion (**Fig. 4d (ii)**), which leads to a broadening of the distribution of the electron hole radii. According to Equation (2), this leads to a distribution of the dipolar-coupling energies $D$ and thus to a broadening of the ELDMR line. In contrast, PLDMR probes triplet states formed via ISC after optical excitation, i.e. without applied voltage. The respective processes are illustrated in **Fig. 4e**. Initially excited donor singlets ($^1LE_D$) can diffuse and undergo ISC to form a molecular triplet ($^3LE_D$) on the donor (**Fig. 4e (i)**). In this case, electron and hole are located on the same molecule resulting in a broad PLDMR spectrum. We measured a PLQY of 6% in oxygen-free m-MTDATA solid films. A significant amount of optically excited singlet states must therefore undergo a non-radiative transition instead of emitting light. Potential non-radiative decay channels are given by non-radiative decay to the singlet ground state or by ISC to the triplet excited state. A PLQY as low as 6% indicates a non-negligible contribution of ISC which explains the population of donor triplets. Alternatively, donor singlets can undergo CT to form a singlet exciplex state ($^1Exc$) (**Fig. 4e (ii)**). The depopulation of donor singlets via both ISC and CT appears to be very efficient since we do not observe any emission from the donor phase in blended m-MTDATA:3TPYMB film (**Fig. 2a**). At low temperatures, the contribution of ISC appears to increase as the broad PLDMR component is more pronounced. This behavior might be explained by a decreased rate of CT, which, according to,[45] is mediated by temperature activated molecular vibrations. Magnetic field effect studies on m-MTDATA:3TPYMB blends report that optically excited exciplex states can diffuse over distances of up to 10 nm within the film, while the electron-hole distance increases during this diffusion process [11,13] (**Fig. 4 e (iii)**). The consequence for PLDMR would be that the dipolar interaction $D$ and the linewidth decrease, yielding the narrow component.

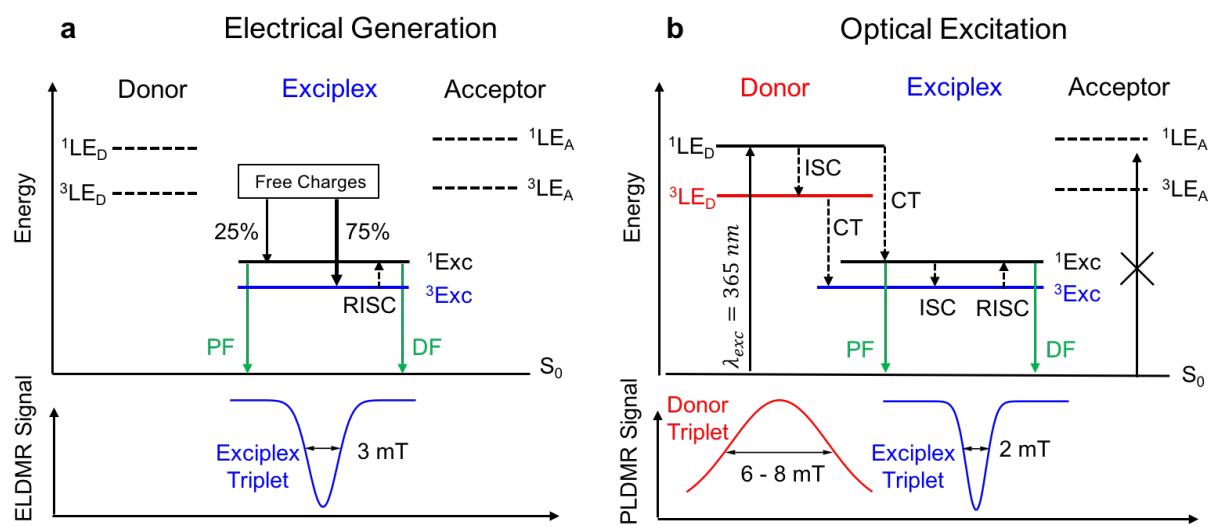

**FIG. 5**. Energy diagrams of excitation pathways. **a)** Excitation of exciplex singlet ($^1Exc$) and triplet ($^3Exc$) via electrical generation in OLED devices. $^1Exc$ can decay directly emitting prompt fluorescence (PF), while RISC between $^3Exc$ and $^1Exc$ enables delayed fluorescence (DF). In ELDMR a single Gaussian line is observed which is attributed to the exciplex triplet. **b)** Population of donor singlets ($^1LE_D$) via optical excitation. Depopulation via CT to exciplex singlet ($^1Exc$) and ISC to donor triplet ($^3LE_D$). Subsequent ISC of $^1Exc$ populates $^3Exc$. These triplets can be upconverted back to singlets, enabling DF. Donor triplets ($^3LE_D$) formed via ISC can populate $^3Exc$ via CT. In PLDMR a broad signal of the donor triplet and a narrow signal of the exciplex triplet is observed.

**Fig. 5** summarizes the excitation pathways of singlet and triplet states for electrical generation and optical excitation. Exact energies of singlet and triplet states for all studied donor:acceptor systems are shown in **Fig. S12**, however the relative positions of the energy levels in **Fig. 5** are representative for all investigated systems. In an electrically driven OLED, free charges are injected to form excitons in the emissive layer. As presented in **Fig. 5a**, the exciplex singlet ($^1$Exc) and triplet ($^3$Exc) are the energetically lowest states within the system, which is why they are preferably populated in a 25:75 ratio. Singlets decay immediately as prompt fluorescence (PF) whereas triplets can undergo thermally activated RISC giving rise to delayed fluorescence (DF). In ELDMR, spin-flip transitions in the exciplex triplet significantly reduce EL intensity by intensifying the non-radiative triplet decay channel. Neither can the emission of molecular singlet excitons in the EL spectrum of an OLED be observed, nor can characteristic signatures of molecular triplets be found in ELDMR. Therefore, triplets of pristine materials ($^3$LE$_D$ or $^3$LE$_A$) are hardly involved in the light generation mechanism of electrically driven devices.

Another scheme can be drawn for the optical excitation of the donor:acceptor blends studied in our work, as illustrated in **Fig. 5b**. Here the UV excitation at 365 nm first generates singlet excitons on donor molecules ($^1$LE$_D$), while the acceptor molecule with an even larger band gap cannot be excited at this wavelength. The singlet exciton of the donor undergoes either ISC to the donor triplet state ($^3$LE$_D$) or CT, which forms the exciplex singlet ($^1$Exc) on a picosecond time scale. Donor triplets subsequently undergo CT to exciplex triplets ($^3$Exc). Exciplex singlets ($^1$Exc) decay radiatively as PF or form exciplex triplets ($^3$Exc) via ISC. Finally, the exciplex triplets are upconverted back to singlet exciplexes via thermally-activated RISC, giving rise to DF. In PLDMR, the exciplex triplets appear as a narrow signal, which is almost identical to the ELDMR signal. A broad component in PLDMR is due to the molecular triplet exciton on the donor molecule, however it's population is strongly temperature dependent. Note that the molecular triplets on the acceptor molecules were not considered in the scheme because they could not be generated with the available excitation energy.

## VI. Conclusions

We applied three different spin-sensitive techniques in order to reveal the role of spin-bearing excited states in the light generation mechanism of TADF-based films and OLEDs. In the case of electrical injection of charge carriers in the donor:acceptor emissive layer, which leads to EL, we found a characteristic signature of the triplet exciplex states ($^3$Exc), both in EL and in electrically detected magnetic resonance. By driving spin-flip transitions within triplet exciplex states the EL intensity changes by up to 1%. This is much higher than what is expected from the Boltzmann statistics at room temperature. From the temperature dependence, we deduced the activation energy of this process, which depends strongly on the donor:acceptor combination and is between 16 and 58 meV. We have assigned this characteristic energy to a singlet-triplet gap and the underlying mechanism of up-conversion to RISC. The underlying TADF mechanism includes a kinetic singlet-triplet exchange based on the g-factor difference, which ensures the conservation of angular momentum, while energy conservation is ensured by a thermal energy supply. As no molecular triplets are observed in the OLED experiments, we suppose they cannot be excited electrically. In the case of optical excitation of donor:acceptor films, the signature of triplet exciplex states ($^3$Exc) is observed in PL and it is similar to the one observed in EL from OLEDs. Additional spectral features from the triplet excitons localized on the donor molecules ($^3$LE$_D$) are also found, but their appearance strongly depends on the material system and on the temperature. Our experiments clearly show that the excited state which is majorly responsible for the occurrence of TADF in donor:acceptor systems is the triplet exciplex state ($^3$Exc). Molecular triplet excitons do not show up at all in efficient OLEDs, but only appear under optical excitation and in some systems mainly at low temperatures. We also emphasize the importance of comparative spin-sensitive, temperature-dependent PL and EL measurements, since the intermediate generation and recombination pathways may differ substantially.

## Methods

The materials m-MTDATA and BPhen were purchased from Sigma-Aldrich. 3TPYMB was purchased from Lumtec (Luminescence Technology Corp.). THCA was supplied by A. Dabuliene and Prof. J. V. Grazulevicius from the Department of Polymer Chemistry and Technology at Kaunas University of Technology, Radvilenu pl. 19, LT-50254 Kaunas, Lithuania. All materials were used as received.

An overview of all studied samples, their preparation and measurements is given in the SI in **Table S1**.

PL samples were prepared by evaporating the emitting layer onto glass substrates. Evaporation rates for all organic materials were 1 Å/s for PL and all other samples. PL and photoexcitation spectra were measured with a calibrated fluorescence spectrometer FLS980-s (Edinburgh Instruments) equipped with continuous broad-spectrum xenon lamp Xe1. Time-resolved PL images were taken with a Streak Camera C5680 (Hamamatsu Photonics) using the second harmonic (400 nm) of MaiTai laser (Spectra-Physics) as an excitation source.

PLDMR was measured either on the same samples as for PL or samples were prepared from solutions of the materials in chlorobenzene. Both sample preparations yielded the same PLDMR results. For solution processing ~100 µl were poured into EPR quartz tubes and the solvent was then evaporated by vacuum pumping. The sample tubes were subsequently sealed under inert helium atmosphere.

PLDMR measurements were done in a modified X-Band spectrometer (Bruker E300, see **Fig. S14**). The sample tube was inserted into an EPR microwave cavity with optical access (Bruker ER4104OR) and an Oxford cryostat (ESR900). Optical excitation was provided by a 365 nm UV LED. The PL was detected by a silicon photodiode placed in front of the cavity behind a 409 nm longpass filter.

All OLED devices were fabricated on indium tin oxide (ITO) covered glass substrates (1 cm$^2$). First, poly(3,4-ethylendioxythiophene):polystyrolsulfonate (PEDOT:PSS, 4083Ai) from Heraeus was spin coated, resulting in a 40 nm thick film. All further device fabrication steps were done inside a nitrogen glovebox to avoid degradation, starting with annealing of the PEDOT:PSS layer for 10 minutes at 130°C. For m-MTDATA:3TPYMB devices, 30 nm donor and 30 nm acceptor were thermally evaporated in an evaporation chamber with an additional 70 nm mixed layer (1:1) in between the pristine material layers. For THCA:BPhen devices, THCA was spin-coated from chlorobenzene solution yielding a layer thickness of 50 nm followed by evaporation of a 40 nm BPhen layer (bilayer device). For m-MTDATA:BPhen both methods were used yielding identical results for the spin-dependent EDMR and ELDMR measurements. The top electrode for all devices was evaporated (5 nm Ca / 120 nm Al), completing OLEDs with 3 mm2 each. Evaporation rates for Ca were 0.3 Å/s and for Al 3 Å/s. For determination of layer thicknesses in spin coated and evaporated materials, a film of the material was scratched with a scalpel and the depth of this scratch was measured with a profilometer (Veeco Dektak 150)

EL spectra were recorded by biasing the OLED with an Agilent 4155C parameter analyzer in constant current mode and coupling the emitted light via light guides to an Acton Spectra SP-2356 spectrometer (Princeton Instruments) or a SPM002 spectrometer (Photon Control).

External quantum efficiencies and luminance were determined by placing an OLED at a distance of 20 mm from a 1 cm$^2$ area Hamamatsu S2281 Si photo detector. The OLED was forward biased via an Agilent 4155C parameter analyzer and the Si photodiode current was collected by the same. Knowledge about the spectral distribution of the OLED emission, the spectral response of the Si-photo detector, and the assumption of a lambertian emitter, allowed determination of the absolute EL photon flux from the OLED and the calculation of the external quantum efficiency and the luminance of the OLED. For temperature dependent measurements the OLED was placed on a peltier element. The construction is shielded by an aluminum housing which is continuously flooded with dry nitrogen in order to prevent degradation by air.

ELDMR measurements were either done in the microwave cavity (bilayer devices of THCA:BPhen and m-MTDATA:BPhen) or devices were placed in contact to a microwave transmission line (co-evaporated devices of m-MTDATA:BPhen and m-MTDATA:3TPYMB) (**see Fig. S15**). This changes slightly the ELDMR intensity due to different coupling of microwave field intensity to the OLED. Signal shape and analysis are however unaffected. For measurements in the cavity a continuous helium flow cryostat and for the stripline measurements a continuous nitrogen flow cryostat provide temperature control and protection of OLEDs from degradation by air. The EL was detected by a silicon photodiode. For all OLEDs constant current (for ELDMR) or constant voltage (for EDMR) forward bias was provided by a source-measure unit (Keithley 237). For all ELDMR, PLDMR and EDMR measurements EL, PL and bias currents were fed to a current-voltage transimpedance amplifier (Femto). The signal change upon resonant microwave irradiation was then detected via a lock-in-amplifier (SR7230) with the on-off modulated microwave as reference.

**Acknowledgements**
N.B. and A.S. acknowledge support by the German Research Foundation, DFG, within the SPP 1601 (SP1563/1-1). S.W. acknowledges DFG FOR 1809 (DY18/12-2). L.K acknowledges H2020-MSCA-ITN-2016 "SEPOMO". V.D. acknowledges the German Research Foundation, DFG, within GRK 2112.

**Author contributions**
N.B., S.W., A.S. and V.D. designed the experiments. N.B., S.W., B.K. and J.G. prepared the devices. N.B., S.W., B.K. and J.G. carried out electro-optical device characterization. N.B., S.W., B.K, J.G and A.S. performed the magnetic resonance measurements. L.K. measured photoexcitation spectra and TRPL. A.S. and V.D. supervised the research project. N.B., S.W. and A.S. evaluated the data and wrote the manuscript together with VD, which all authors discussed and commented on.

**Competing Financial Interests statement**
The authors declare no competing financial interests

# Supporting Information for

# Optically and electrically excited intermediate electronic states in donor:acceptor based OLEDs

Nikolai Bunzmann[1], Sebastian Weissenseel[1], Liudmila Kudriashova[1], Jeannine Gruene[1], Benjamin Krugmann[1], Juozas Vidas Grazulevicius[2], Andreas Sperlich[1*] and Vladimir Dyakonov[1,3]

[1] Experimental Physics VI, Julius Maximilian University of Würzburg, 97074 Würzburg, Germany
[2] Department of Polymer Chemistry and Technology at Kaunas University of Technology, Radvilenu pl. 19, LT-50254 Kaunas, Lithuania

(Dated: January 10, 2020)

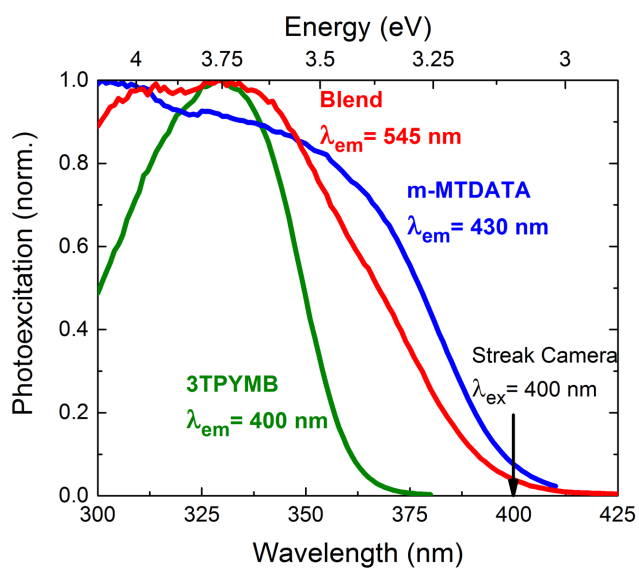

**FIG. S1**. Photoexcitation spectra of solid films of m-MTDATA, 3TPYMB and their blend. Excitation with a 400 nm laser in the streak camera measurements, or with a 365 nm UV LED in PLDMR will only excite m-MTDATA, but not 3TPYMB.

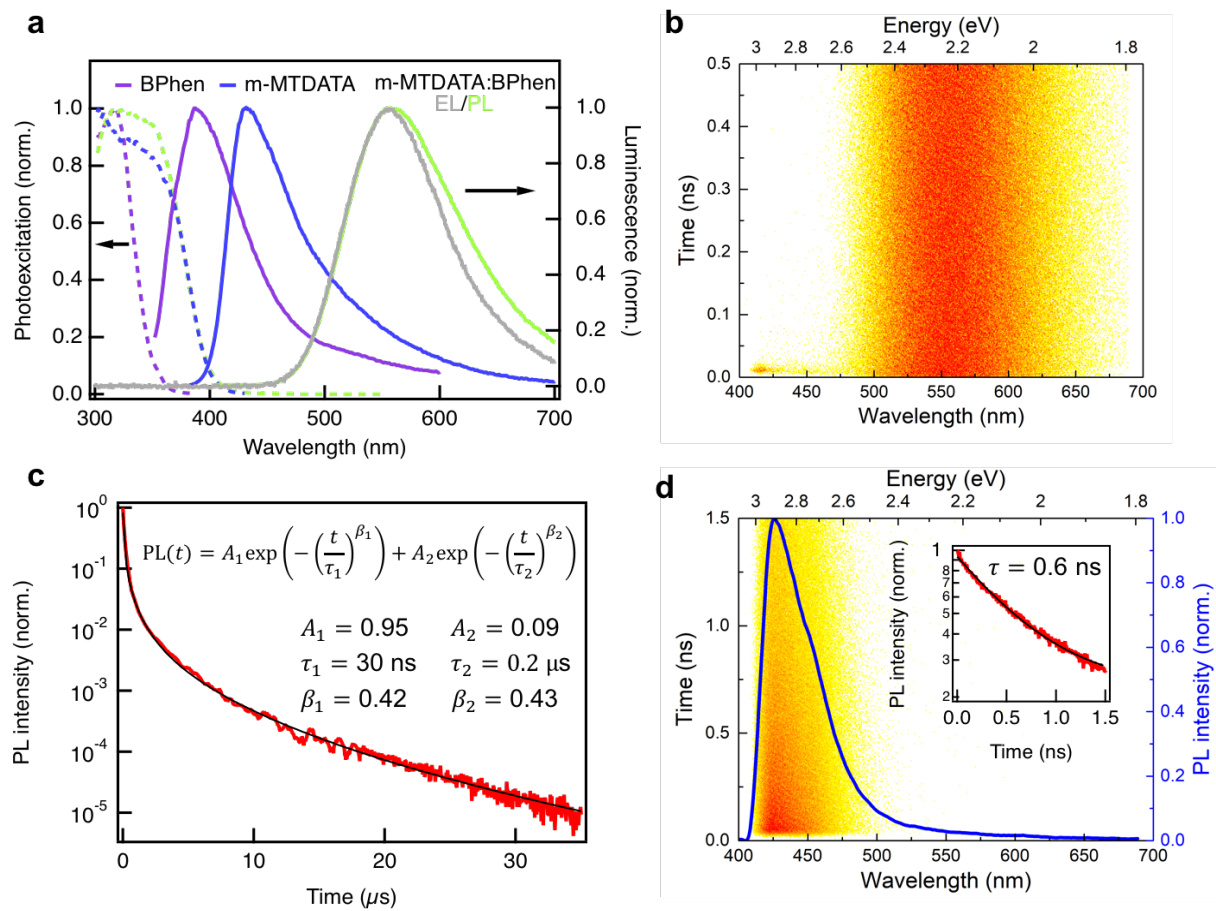

**FIG. S2**. Photophysics of m-MTDATA:BPhen. **a)** PE and PL spectra of solid films of BPhen, m-MTDATA and their blend together with an EL spectrum of an OLED based on m-MTDATA:BPhen. **b)** Streak camera image of PL from an m-MTDATA:BPhen blended solid film for excitation at $\lambda_{exc} = 400$ nm. **c)** Transient PL decay of an m-MTDATA:BPhen blended solid film. A sum of two stretched exponential decays was used to fit the curve and the resulting parameters are listed in the inset. Note that the characteristic lifetimes obtained from the stretched exponential decay do not necessarily represent the lifetimes of prompt and delayed fluorescence. **d)** Streak camera image of PL from a solid film of pristine m-MTDATA. Inset: PL decay curve of this emission. A lifetime of 0.6 ns is obtained from fitting the transient with a single exponential decay.

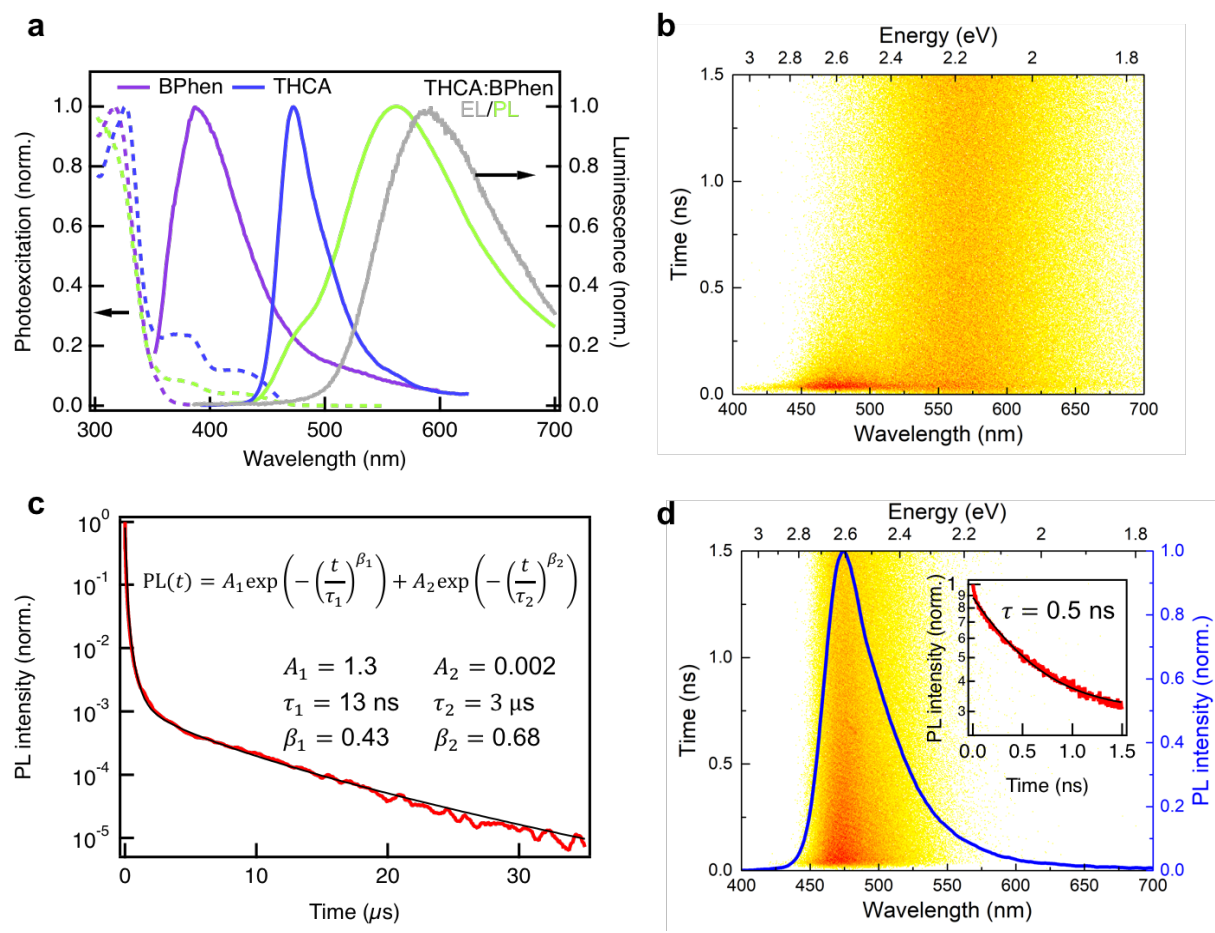

**FIG. S3**. Photophysics of THCA:BPhen. **a)** PE and PL spectra of solid films of BPhen, THCA and their blend together with an EL spectrum of a bilayer OLED based on THCA:BPhen. **b)** Streak camera image of PL from a THCA:BPhen blended solid film for excitation at $\lambda_{exc}$ = 400 nm. **c)** Transient PL decay of a THCA:BPhen blended solid film. A sum of two stretched exponential decays was used to fit the curve and the resulting parameters are listed in the inset. Note that the characteristic lifetimes obtained from the stretched exponential decay do not necessarily represent the lifetimes of prompt and delayed fluorescence. **d)** Streak camera image of PL from a solid film of pristine THCA. Inset: PL decay curve of this emission. A lifetime of 0.5 ns is obtained from fitting the transient with a single exponential decay.

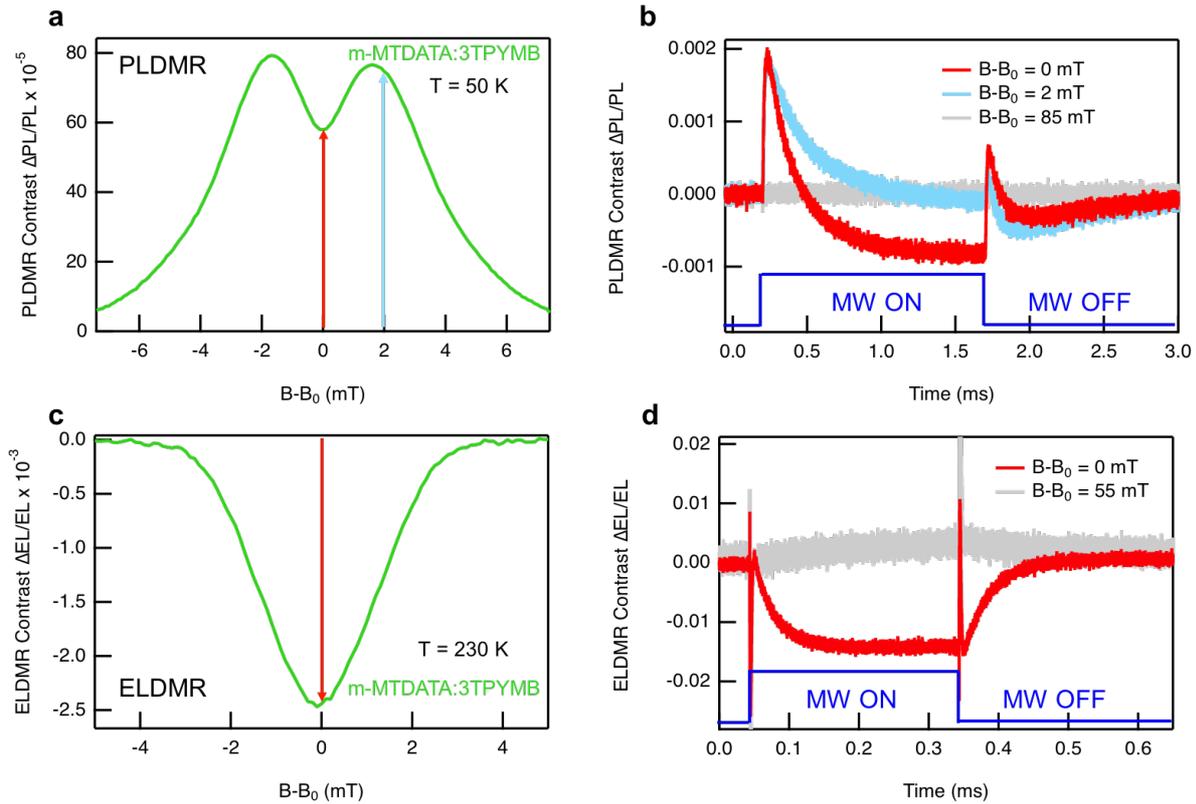

**FIG. S4**. Comparison of lock-in and directly measured PLDMR and ELDMR signals **a)** Lock-in detected PLDMR signal of an m-MTDATA:3TPYMB blend at 50 K. **b)** Transient measurement of the microwave induced change of PL at a constant external magnetic field. The time-dependent change of PL is measured at different magnetic field positions of the PLDMR spectrum (Graph **a**), as indicated by the arrows. **c)** Lock-in detected ELDMR signal of an OLED based on m-MTDATA:3TPYMB at 230 K. **d)** Transient measurement of the microwave induced change of EL at a constant external magnetic field. The time-dependent change of EL is measured at a magnetic field position in resonance which is indicated by an arrow in the graph of the lock-in detected ELDMR signal (Graph **c**). The off-resonance measurements (gray), which determines the baseline, and the on-off microwave modulation sequences (blue) are also shown.

In order to determine the sign of PLDMR and ELDMR signals we used an oscilloscope to monitor the time dependent microwave induced change in PL/EL during microwave on-off modulation at a constant magnetic field.

For PLDMR, transients at three different magnetic field positions are measured. For the first measurement the magnetic field is set to $B-B_0 = 0$ mT where narrow and broad component are superimposed in the lock-in detected signal (**Fig. S4a**, red arrow). The time trace measured with the oscilloscope exhibits a rapid increase of PL, when resonant microwaves are switched on, which decays over time afterwards (**Fig. S4b**, red curve). The intensity does not decay back to the initial PL intensity but there is an additional negative offset. In a second measurement the magnetic field is set to $B-B_0 = 2$ mT where only the broad signal is in resonance (**Fig. S4a**, blue arrow). Here the time trace measured with the oscilloscope exhibits a rapid increase of PL but in this case the intensity decays back to its initial intensity (**Fig. S4b**, blue curve). Finally, a third measurement at $B-B_0 = 85$ mT shows that there is no off-resonance microwave induced change of PL (**Fig. S4b**, grey curve). From the difference between time traces at $B-B_0 = 0$ mT and $B-B_0 = 2$ mT we conclude that the rapid increase is attributed to the broad lock-in detected component and the negative offset to the narrow component. Furthermore, these measurements reveal that the broad component corresponds to an increase of PL intensity and the narrow component to a decrease. Based on these findings we are now able to set the sign of the superimposed signal components in lock-in detected PLDMR measurements with certainty: narrow components exhibit a negative sign while broad components exhibit a positive sign.

For ELDMR, transients at two different magnetic field positions are measured. For the first measurement the magnetic field is set to $B-B_0 = 0$ mT where the resonance condition is fulfilled (**Fig. S4c**, red arrow). The corresponding time trace measured with the oscilloscope exhibits a decay of EL intensity when resonant microwaves are switched on (**Fig. S4d**, red curve). This decay saturates at a level below the initial EL intensity.

From this measurement we conclude that the microwave induce change of EL in resonance corresponds to a decrease of EL intensity. Therefore, we set the sign in lock-in detected ELDMR measurements to negative.

The effect of static magnetic field on EL intensity (magneto-electroluminesence, MEL) (Fig. S5) confirms the built-up of spin polarisation of triplet exciplex state in magnetic field due to different lifetimes of the triplet Zeeman sublevels $m_S$=0, +/-1.

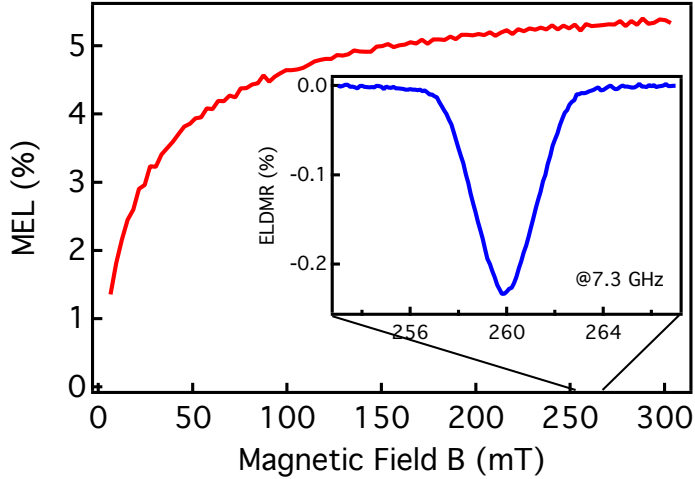

**FIG. S5.** Magnetic field effect on electroluminescence (MEL) in an OLED based on m-MTDATA:3TPYMB at T= 220 K. EL enhancement is measured directly. Inset: Separately measured ELDMR on the same device with 7.3 GHz on/off modulated microwaves and lock-in phase-sensitive detection. The sign of ELDMR is directly determined from the transient EL measurements as in **Fig. S4d**.

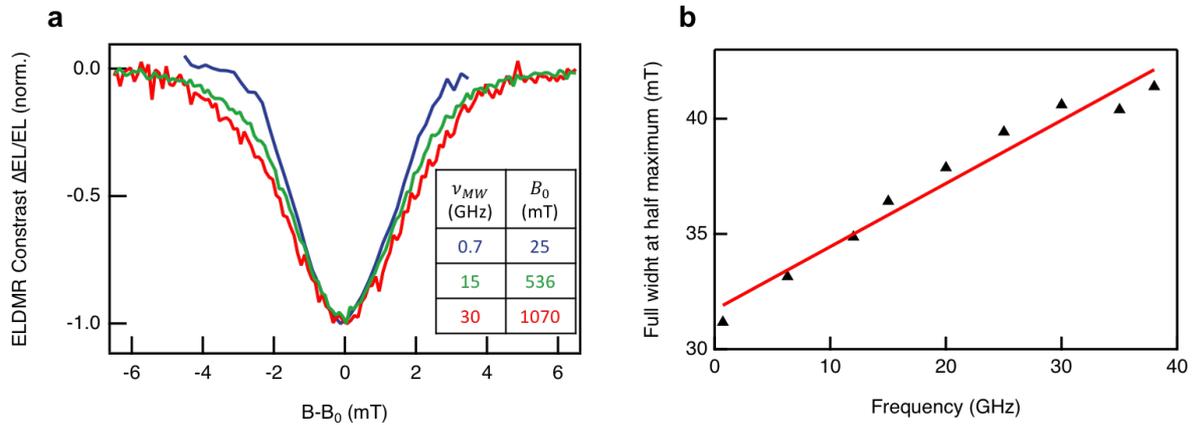

**FIG. S6. a)** Normalized ELDMR spectra of m-MTDATA:3TPYMB based OLEDs at different microwave frequencies $\nu_{MW}$ and magnetic fields $B_0$. The magnetic field axis is shifted, such that resonance peaks at $B_0$ are centred around $B - B_0$. The temperature for all measurements was T = 230 K. **b)** Frequency dependence of the full width at half maximum of ELDMR spectra.

The microwave frequency dependence of the linewidth of ELDMR spectra can be used to estimate the difference $\Delta g$ in $g$-factors of electron and hole forming the exciplex state. The following considerations explain the theoretical background of this method.

Under magnetic resonance conditions the following equation is fulfilled:

$$h\nu_{MW} = g\mu_B B \Delta m_s \qquad (S1)$$

Here $\nu_{MW}$ is the microwave frequency, $h$ is the Planck constant, $g$ is the $g$-factor of the spin, $\mu_B$ is the Bohr magneton, $B$ is the magnetic field value and $\Delta m_s = 1$ is the allowed change of the magnetic quantum number. For two independent spins with $g$-factors $g_a$ and $g_b$ (in the case of an exciplex state these two spins correspond to electron and hole spin located on separate molecules) the difference $\Delta B$ of the resonance positions of the two spins on the magnetic field axis is given by:

$$\Delta B = \left(\frac{1}{g_a} - \frac{1}{g_b}\right)\frac{h\nu_{MW}}{\mu_B} \quad (S2)$$

According to equation S2 the magnetic field splitting $\Delta B$ depends linearly on the microwave frequency $\nu_{MW}$ if $\Delta g = |g_a - g_b| > 0$. At low frequencies, $\Delta B$ is negligible and the resonance curves of the two spins overlap. In this case the linewidth of the overall EPR signal is dominated by broadening mechanisms such as dipolar interaction $D$ and unresolved hyperfine interactions. For higher microwave frequencies the increase of $\Delta B$ causes an additional broadening of the EPR linewidth as shown in Figure S6.

The resonant transmission line approach we use to measure ELDMR spectra (see experimental section and **Fig. S17**) allows us to measure at different microwave frequencies and probe the above described behavior. We were able to measure ELDMR spectra in a frequency range between 0.7 and 38 GHz. Exemplary spectra for 0.7, 15 and 30 GHz are shown in **Fig. S6a**. Here, an increase of the linewidth for higher frequencies can be observed. Evaluation of the frequency dependence of the full width at half maximum for all ELDMR spectra in the frequency range between 0.7 and 38 GHz shows a roughly linear dependence as depicted in **Fig. S6b**. This observation is in line with equation S2 and indicates $\Delta g > 0$.

An upper boundary value for $\Delta g$ can be extracted from a simulation of the ELDMR spectra within the whole frequency range. For this purpose we used the software package EasySpin,[1] which is commonly used to extract parameters from EPR spectra. Here, a system of two coupled spins with different $g$-factors was assumed. Every spectrum in the accessible frequency range was included in a global fit. We use $\overline{g} = (g_a + g_b)/2$ as the mean $g$-factor and $\Delta g = |g_a - g_b|$ as the difference in fixed, scalar $g$-factors. Alternatively, anisotropy ($g$-tensor) or a possible distribution of exciplexes with slightly different $g$-factors ($g$-strain) would also result in a magnetic field-dependent broadening. This can be caused by the flexibility of the involved molecules, as well as by the undefined geometry of interacting donor and acceptor molecules hosting an exciplex. These possibilities cannot be discerned with this data set alone, as no actual splitting of the resonance spectrum is observed that would directly indicate $\Delta g$ as described in equation S2. The choice for this preliminary analysis to fit with fixed $g$-factors and leaving out $g$ anisotropy and $g$ strain still yields an upper limit for the value of $\Delta g$. Yet, this limit embraces any actual $g$-tensor strain or anisotropy. From our dataset we obtain $\Delta g < 9.2 \cdot 10^{-4}$. The value of the mean $g$-factor $\overline{g} = 2.0007$ is close to the $g$-factor of a free electron and typical for polarons in organic semiconductors. We are not able to assign separate values to the electron and hole but rather can only determine the difference between them.

The fit is performed using EasySpin - a software toolbox based on Matlab, which simulates EPR spectra of spin systems. The starting parameters for the fit are defined as follows:

```
Sys.S=[1/2 1/2]; %specifies the system as two spin ½ carrying particles.
Sys.g = [2.00118 ; 2.00021]; %sets the initial values for the g-factors of the two spins.
Sys.D=[27]; %sets the initial value for the dipolar interaction to 27 MHz.
Sys.lw=4; %sets the initial value for the Gaussian broadening of the linewidth to 4 mT.
```

First, the spin system is defined as two spin ½ particles with corresponding $g$-factors, which, in our case, represent electron and hole forming an exciplex state. Furthermore, a dipolar coupling $D$ between the two spins is defined. $D$ basically contains information about the distance of the two spins as described in the main part of this work. An initial linewidth for the EPR spectrum has to be set to account for broadening mechanisms such as unresolved hyperfine coupling. On the one hand, EasySpin allows to *simulate* a theoretical EPR spectrum of such a spin system if the user provides the values for the $g$-factors, the dipolar coupling and the initial linewidth as well as the frequency at which the EPR experiment is performed. On the other hand, EasySpin can be used to *fit* experimental data, where the parameters of the spin system are optimized to reproduce the measured signal. For a global fit this optimization is performed for several spectra simultaneously. General Information on EasySpin can be found in [1] or at easyspin.org/easyspin/documentation/

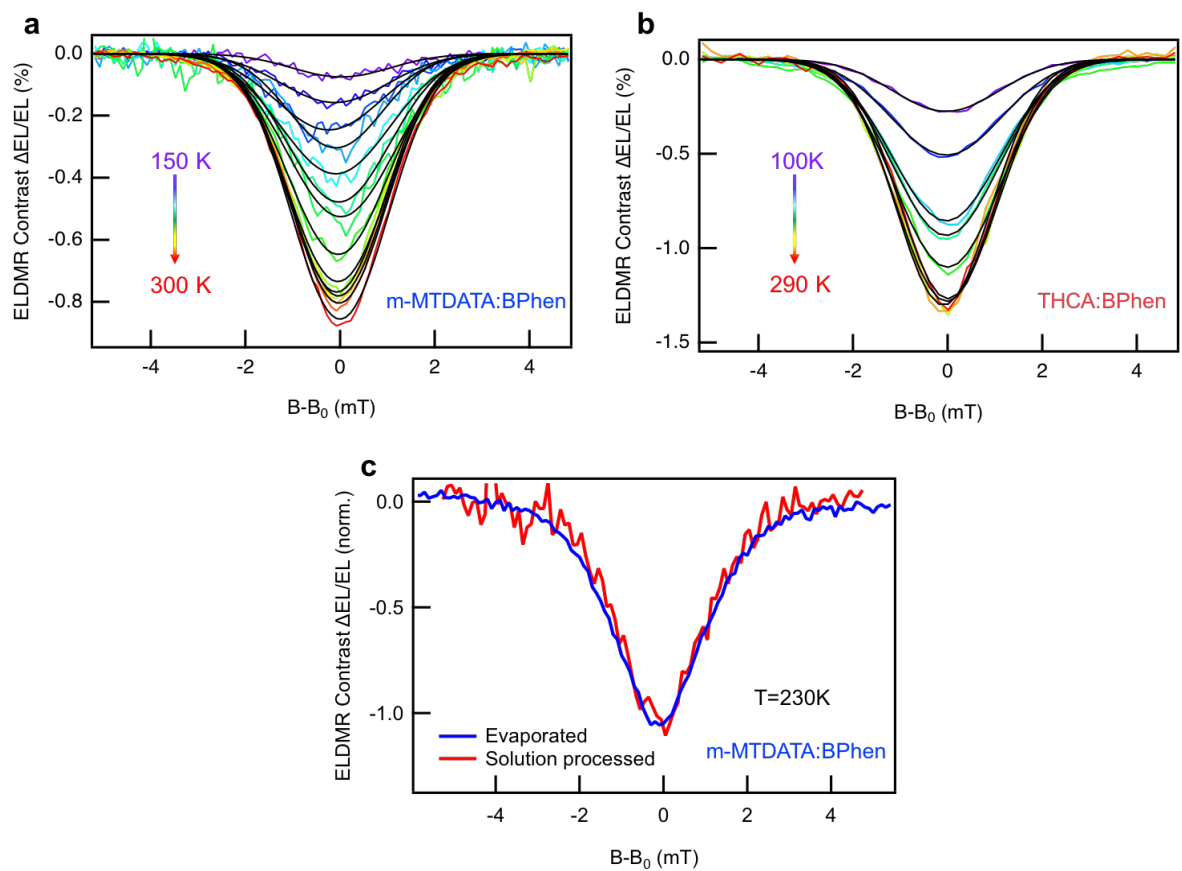

**FIG. S7**. Temperature-dependent ELDMR of **a)** m-MTDATA:BPhen and **b)** THCA:BPhen together with Gaussian fits (black lines). For both material combinations the signal decreases with decreasing temperature, which is in agreement with TADF behaviour. **c)** Normalized ELDMR spectra measured on m-MTDATA:BPhen OLEDs based either on evaporated donor and acceptor (blue), or solution processed donor and evaporated acceptor. Spectra are nearly identical showing that the ELDMR signal does not depend on the preparation method.

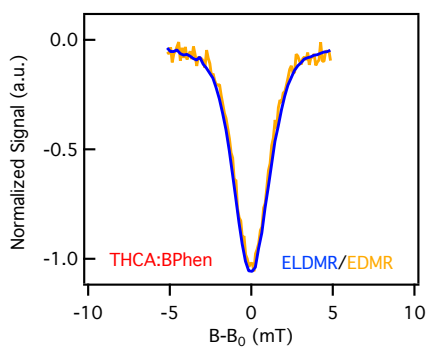

**FIG. S8**. Normalized ELDMR and EDMR spectra for THCA:BPhen. Both methods yield signals with identical shape, therefore originating from the same effect. We assume the enhanced recombination of exciplexes in resonance (ELDMR) induces a change of the current (EDMR).

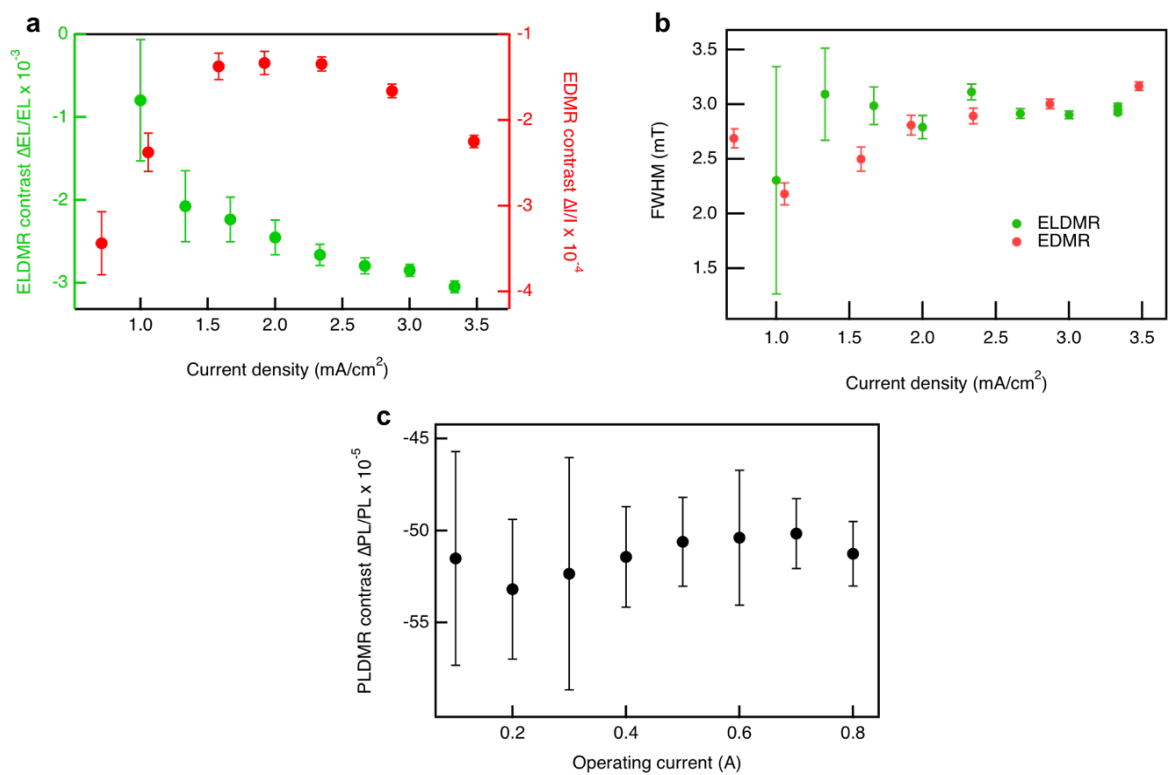

**FIG. S9**. Dependence of the signal intensities of magnetic resonance measurements on excitation power for m-MTDATA:3TPYMB. **a)** Dependence of ELDMR and EDMR signal intensities on the current density in the OLED at T=220 K. **b)** Dependence of the full width at half maximum (FWHM) in EDMR and ELDMR on the current density in an OLED at T=220 K. **c)** Dependence of PLDMR signal intensities on the excitation power at T=RT. A UV LED is used for optical excitation in PLDMR measurements. Note, the power output of the UV LED is proportional to the operating current that is why the x-axis can be treated as excitation power.

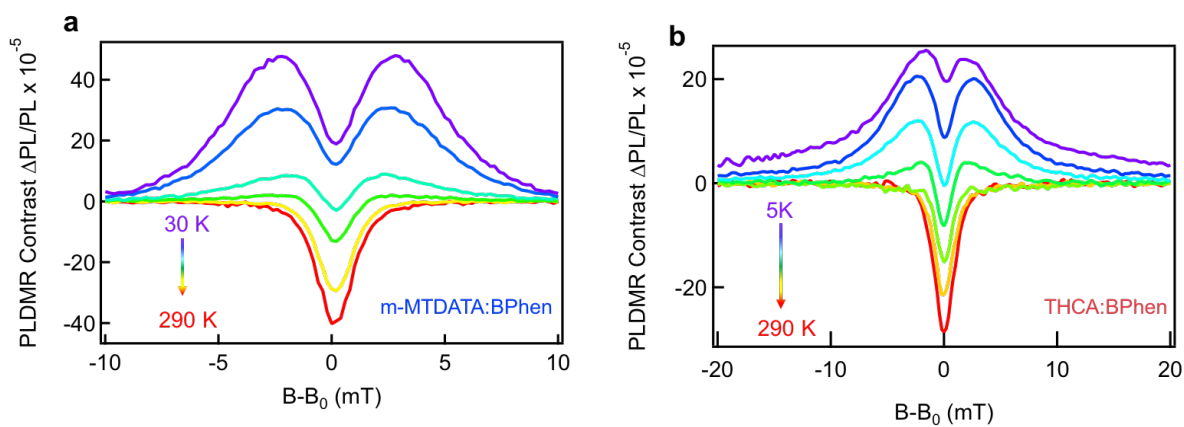

**FIG. S10**. Temperature dependent PLDMR spectra for the donor:acceptor blends **a)** m-MTDATA:BPhen and **b)** THCA:BPhen. In both molecular blends, the PLDMR spectrum consists of a narrow signal which dominates at room temperature and an additional broad signal which is more pronounced at low temperatures. The narrow (negative) signal is assigned to the exciplex triplet and the broad (positive) signal to the local triplet exciton of the respective donor molecule.

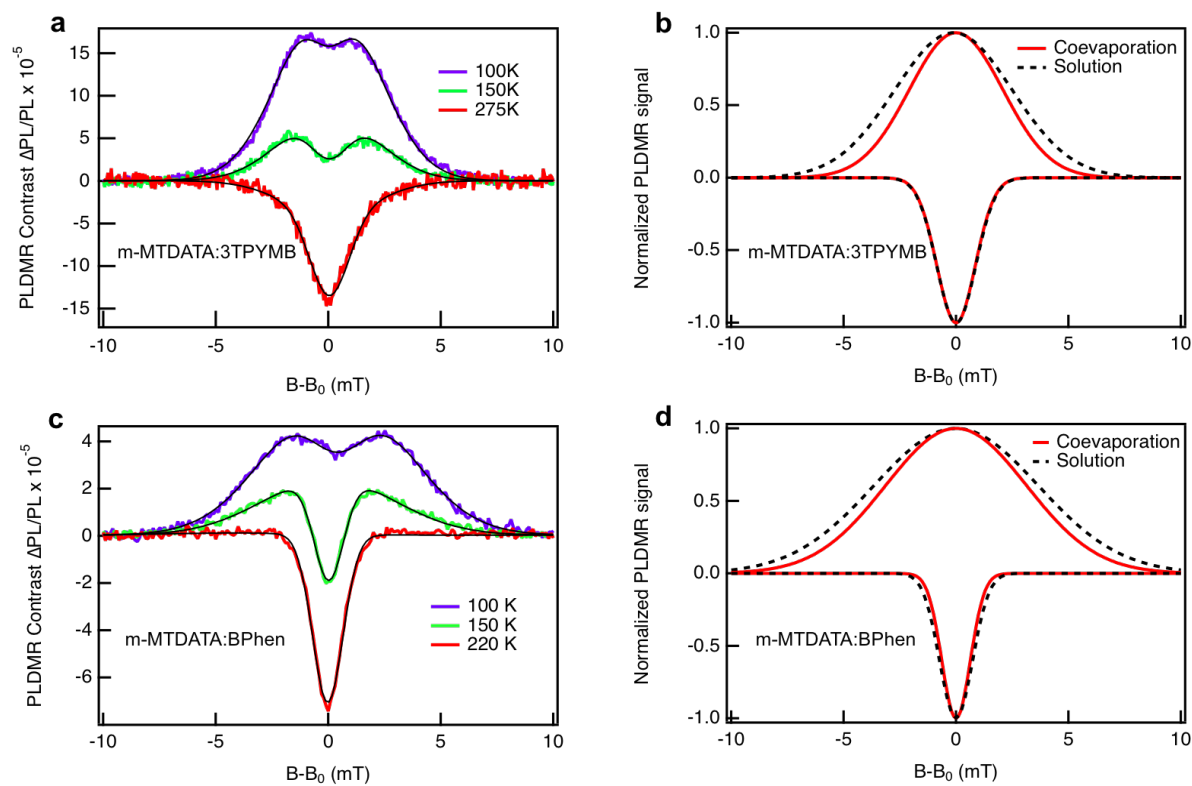

**FIG. S11**. **a)** Temperature dependent PLDMR measurements on a coevaporated m-MTDATA:3TPYMB sample. The signal shape and temperature behaviour are very similar to those measured on solution processed solid films shown in **Fig. 4b** in the main text. **b)** Normalized fit components obtained for solution processed solid films and evaporated m-MTDATA:3TPYMB samples. **c)** Temperature dependent PLDMR measurements on a coevaporated m-MTDATA:BPhen sample. **d)** Normalized fit components obtained on solution processed solid films and evaporated m-MTDATA:BPhen samples. In both molecular systems, the narrow components (exciplex states) are identical, whereas the broad components are slightly narrower in the vacuum processed solid films, which can be attributed to a larger radius of molecular triplet excitons in the latter.

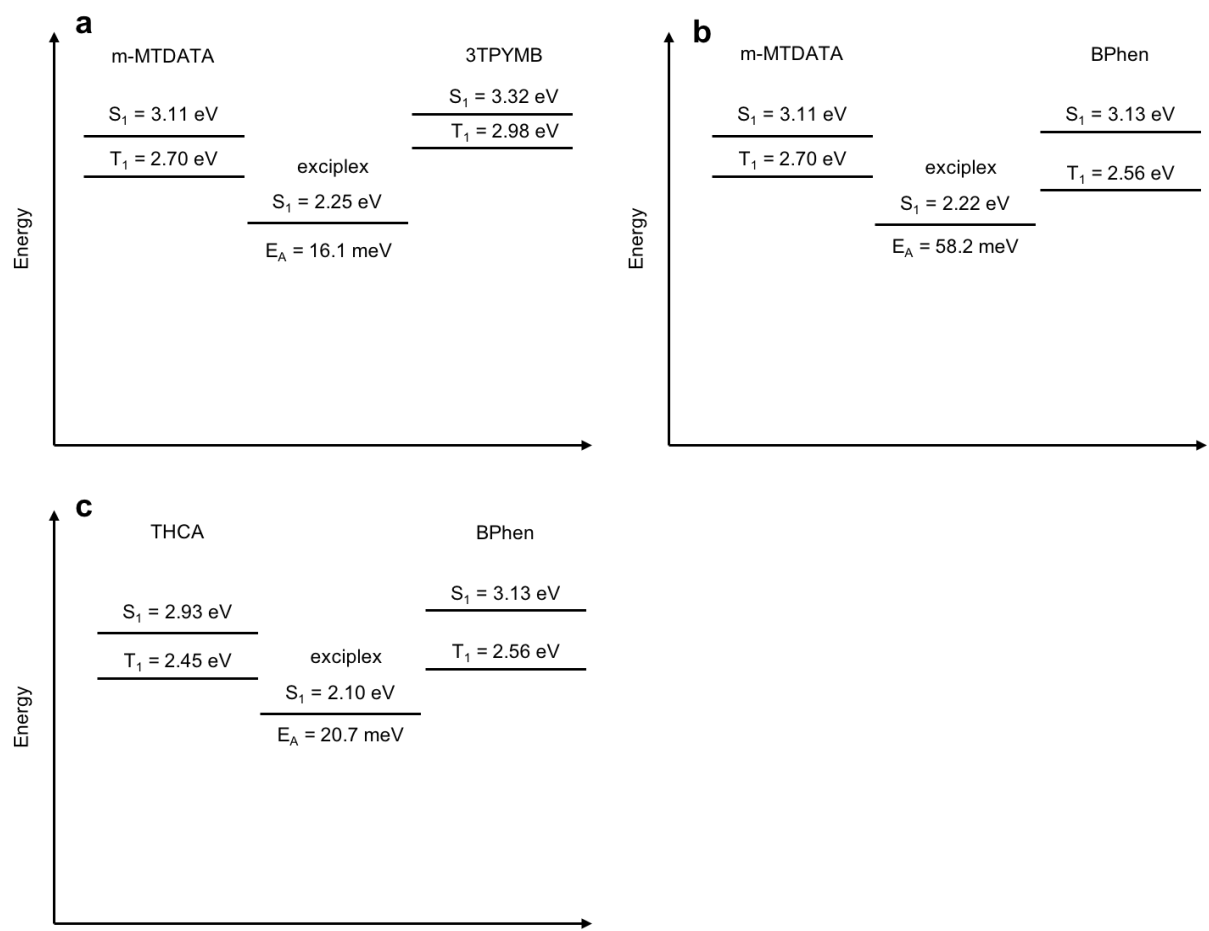

**FIG. S12**. Energy diagrams showing singlet and triplet energies of individual donor and acceptor molecules as well as of the exciplex state formed at the respective interfaces. **a)** m-MTDATA:3TPYMB **b)** m-MTDATA:BPhen **c)** THCA:BPhen. Values for singlet and triplet states of m-MTDATA and 3TPYMB are taken from [2], for BPhen from [3] and for THCA from [4]. Values for the exciplex states were calculated from peaks of PL spectra for each donor:acceptor combination. (**Fig. 1**, **Fig. S2, Fig. S3**) The activation energies are derived from the Arrhenius plot in **Fig. 3c**.

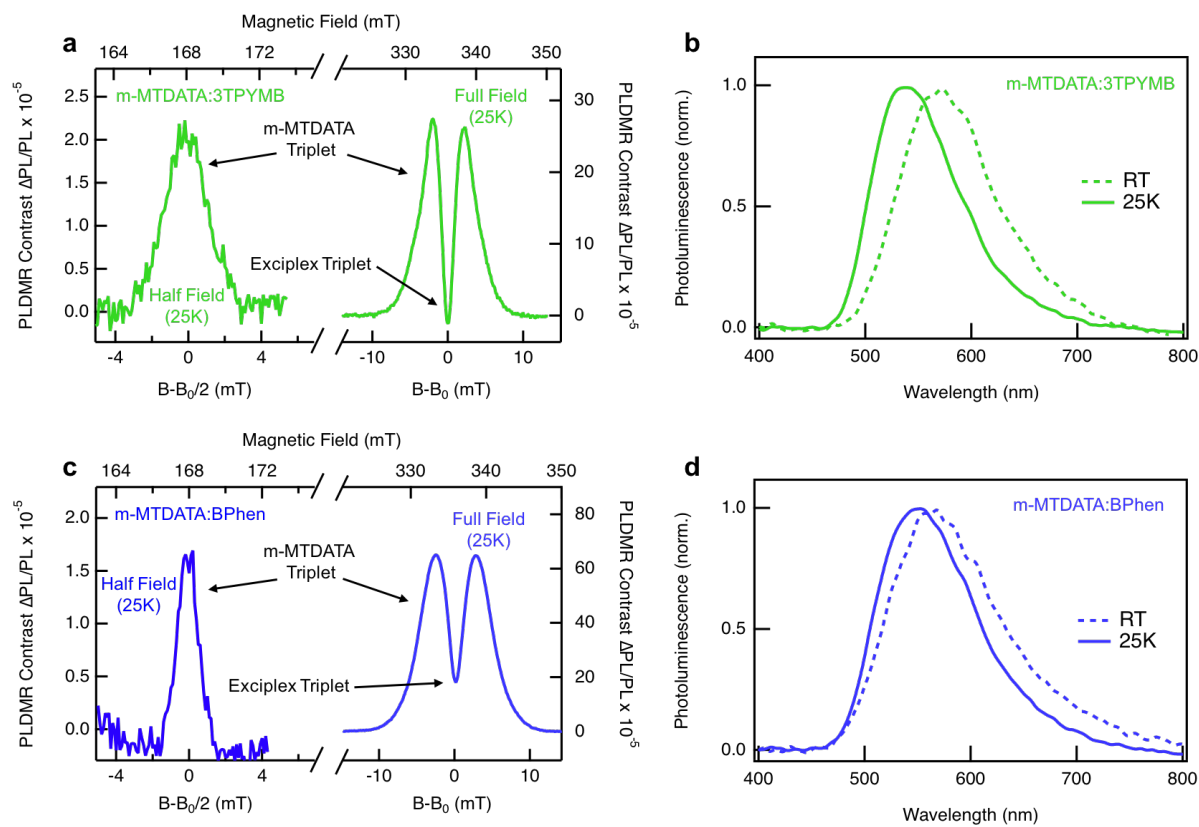

**FIG. S13**. Half-field and full-field PLDMR signals in **a)** m-MTDATA:3TPYMB and **c)** m-MTDATA:BPhen blends. T=25 K. PL spectra from **b)** m-MTDATA:3TPYMB and **d)** m-MTDATA:BPhen blends recorded at T=25 K (solid lines) and RT (dashed lines). In this temperature range, only PL from the exciplex singlet state is visible exhibiting a small blue shift.

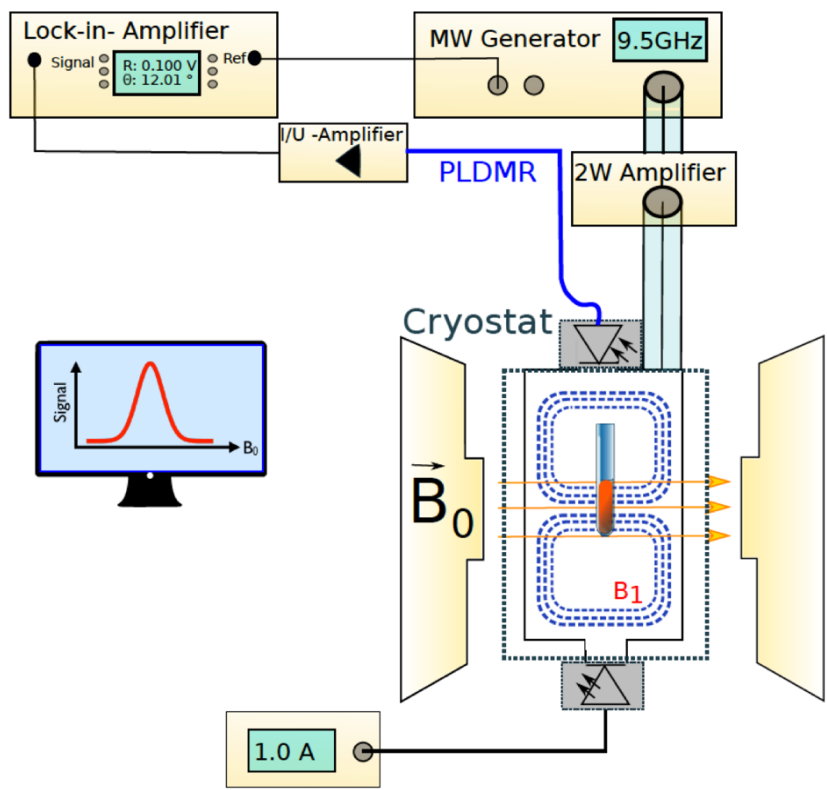

**FIG. S14**. Scheme of PLDMR setup.

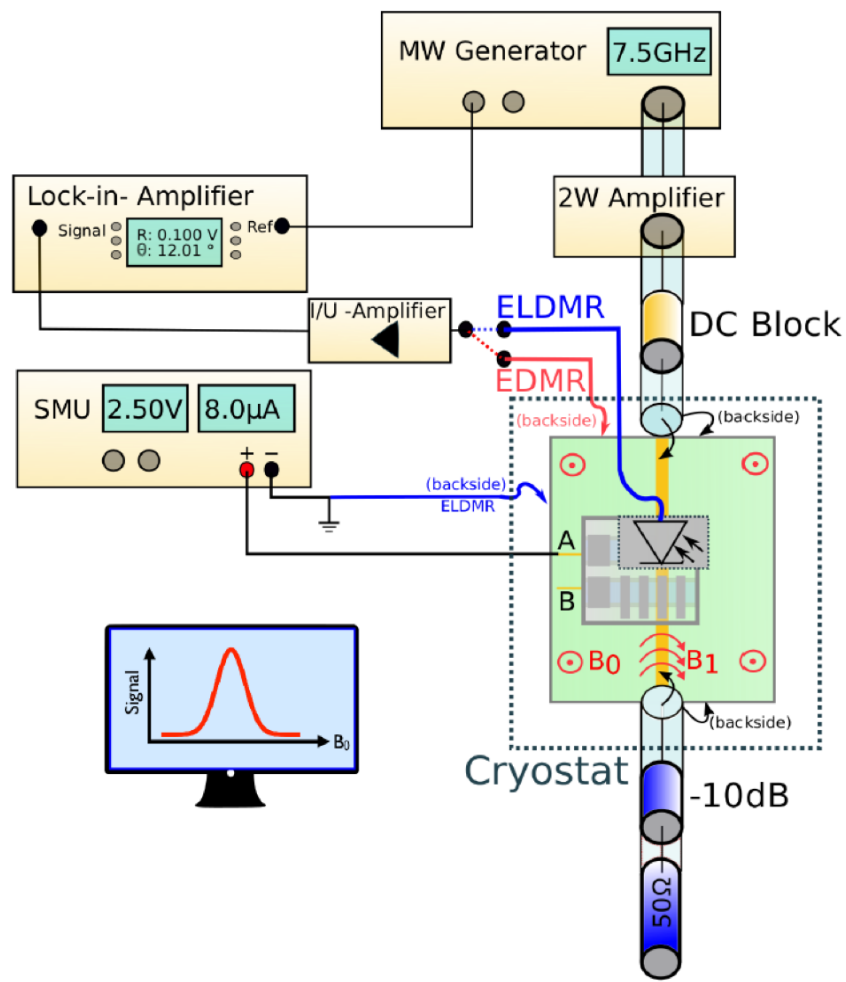

**FIG. S15**. Scheme of ELDMR setup.

|  |  | **Applied method and figure number in the main manuscript and SI**. | | | | | | |
|---|---|---|---|---|---|---|---|---|
| **Material** | **Fabrication** | PL | trPL | PLE | EL | ELDMR | EDMR | PLDMR |
| m-MTDATA | evaporated | 1, S2 | 2, S2 | S1, S2 | | | | |
|  | solution processed | | | | | | | S15 |
| 3TPYMB | evaporated | 1 | not excitable[B)] | S1 | | | | |
|  | solution processed | | | | | | | S15 |
| BPhen | evaporated | S2, S3 | | S2, S3 | | | | not excitable[A)] |
|  | solution processed | | | | | | | |
| THCA | evaporated | S3 | S3 | S3 | | | | |
|  | solution processed | | | | | | | no signal |
| m-MTDATA: 3TPYMB | co-evaporated | 1 | 2 | S1 | 1 | 3,4,S4, S5,S6,S9 | 3, S6, S9 | S11 |
|  | solution processed | S13 | | | | | | 4,S4,S9,S11, S13,S15 |
| m-MTDATA: BPhen | co-evaporated | S2 | S2 | S2 | not shown [C)] | S7 | not shown [C)] | S11 |
|  | solution processed | S13 | | | | | | 4,S10,S11,S13 |
|  | Bilayer: m-MTDATA from solution, BPhen evaporated on top | | | | S2 | 4, S7 | not shown [C)] | |
| THCA:BPhen | co-evaporated | S3 | S3 | S3 | | | | |
|  | solution processed | | | | | | | 4, S10 |
|  | Bilayer: THCA from solution, BPhen evaporated on top | | | | S3 | 4, S7, S8 | S8 | |
| PEDOT:PSS | solution processed | | | | | | 3 | |

[A)] BPhen: PLDMR; not excitable with 365 nm UV LED
[B)] 3TPYMB, BPhen: not excitable with 400 nm laser of streak camera setup
[C)] Identical EL, ELDMR and EDMR spectra for co-evaporated and bilayer OLEDs

**TABLE S1**. Overview of sample preparation and applied methods.

**EQE estimate for m-MTDATA:3TPYMB**

A total photoluminescence quantum yield (PLQY) of 0.45 was measured for an oxygen-free m-MTDATA:3TPYMB solid film, whereas the PLQY of the same film fully quenched by oxygen was 0.04. Then the quantum efficiency of the prompt component is $\varphi_{\text{pr}} = 0.04$ and of the delayed component is $\varphi_{\text{TADF}} = 0.45 - 0.04 = 0.41$. Now, under the assumption that phosphorescence is not present at room temperature, the quantum efficiencies of ISC, RISC, and nonradiative triplet decay can be expressed via $\varphi_{\text{pr}}$ and $\varphi_{\text{TADF}}$: $\varphi_{\text{ISC}} = 1 - \varphi_{\text{pr}} = 1 - 0.04 = 0.96$, $\varphi_{\text{RISC}} = \varphi_{\text{TADF}} = 0.41$, and $\varphi_{\text{nr}} = \varphi_{\text{ISC}} - \varphi_{\text{TADF}} = 0.96 - 0.41 = 0.55$. With that, the maximum internal electroluminescence efficiency of the emitter in an OLED can be expressed as [5]:

$$\Phi_{\text{EL,int}} = \eta_S \varphi_{\text{pr}} + \eta_S \varphi_{\text{ISC}} \varphi_{\text{RISC}} + \eta_T \varphi_{\text{RISC}};$$

$$\Phi_{\text{EL,int}} = 0.25 \cdot 0.04 + 0.25 \cdot 0.96 \cdot 0.41 + 0.75 \cdot 0.41 \approx 0.416;$$

Here $\eta_S$ and $\eta_T$ are the portions of singlets and triplets produced via electrical injection (0.25 and 0.75, respectively). We obtained $\Phi_{\text{EL,int}} = 41.6\%$, which results in the estimation of $\text{EQE}_{\text{max}} = 8.3\%$ in devices with an assumed 20% light outcoupling.